\pdfoutput=1
\documentclass[11pt,twoside,a4paper,cmspaper,final,collab]{cms-tdr}
\def\svnVersion{3637623}\def\svnDate{2026/01/14}\def\cmsCernNoTag{CERN-EP-2025-013}\def\cmsCernDate{\today}\def\cmsMessage{Published in the Journal of High Energy Physics as \href{http://dx.doi.org/10.1007/JHEP01(2026)047}{\doi{10.1007/JHEP01(2026)047}.}}
\begin{document}\cmsNoteHeader{SMP-22-017}

\newlength\cmsTabSkip\setlength{\cmsTabSkip}{1ex}

\newcommand{\acc}{\ensuremath{\,\text{(acceptance)}}\xspace}
\newcommand{\PDF}{\ensuremath{\,\text{(PDF)}}\xspace}
\newcommand{\PDFns}{\ensuremath{\text{(PDF)}}\xspace}
\newcommand{\scale}{\ensuremath{\,\text{(scale)}}\xspace}
\newcommand{\scalens}{\ensuremath{\text{(scale)}}\xspace}
\newcommand{\intns}{\ensuremath{\text{(int)}}\xspace}
\newcommand{\systns}{\ensuremath{\text{(syst)}}\xspace}
\newcommand{\statns}{\ensuremath{\text{(stat)}}\xspace}
\newcommand{\lumns}{\ensuremath{\text{(lumi)}}\xspace}
\newcommand{\pp}{\ensuremath{\Pp\Pp}\xspace}
\newcommand{\DY}{\ensuremath{\PZ/\PGg^\ast}\xspace}
\newcommand{\VV}{\ensuremath{\PV\PV}\xspace}
\newcommand{\mmumu}{\ensuremath{m_{\PGm\PGm}}\xspace}
\newcommand{\sigeta}{\ensuremath{\eta}\xspace}
\newcommand{\abseta}{\ensuremath{\abs{\eta}}\xspace}
\newcommand{\sigfid}{\ensuremath{\sigma_{\text{fid}}}\xspace}
\newcommand{\sigtot}{\ensuremath{\sigma_{\text{tot}}}\xspace}
\newcommand{\BR}{\ensuremath{\mathcal{B}}\xspace}
\newcommand{\MpN}{\ensuremath{\PGmp\PGn}\xspace}
\newcommand{\MmN}{\ensuremath{\PGmm\PAGn}\xspace}
\newcommand{\qt}{\ensuremath{q_{\mathrm{T}}}\xspace}
\newcommand{\DYTurbo}{\ensuremath{\textsc{DYTurbo}}\xspace}
\newcommand{\WoverZ}{\ensuremath{\PW\!/\PZ}\xspace}
\newcommand{\WpoverZ}{\ensuremath{\PWp\!/\PZ}\xspace}
\newcommand{\WmoverZ}{\ensuremath{\PWm\!/\PZ}\xspace}
\newcommand{\WpmoverZ}{\ensuremath{\PWpm\!/\PZ}\xspace}
\newcommand{\WpoverWm}{\ensuremath{\PWp\!/\PWm}\xspace}
\newcommand{\WptoMpN}{\ensuremath{\PWp\!\to\MpN}\xspace}
\newcommand{\WmtoMmN}{\ensuremath{\PWm\!\to\MmN}\xspace}
\newcommand{\ZtoMM}{\ensuremath{\PZ\to\MM}\xspace}
\newcommand{\Rfid}{\ensuremath{R_{\text{fid}}}\xspace}
\newcommand{\Rtot}{\ensuremath{R_{\text{tot}}}\xspace}
\newcommand{\efficiency}[1]{\ensuremath{\epsilon_{\text{#1}}}\xspace}

\cmsNoteHeader{SMP-22-017}

\title{Measurements of the inclusive \texorpdfstring{\PW}{W} and \texorpdfstring{\PZ}{Z} boson production cross sections and their ratios in proton-proton collisions at \texorpdfstring{$\sqrt{s}=13.6\TeV$}{sqrt(s) = 13.6 TeV}}

\date{\today}

\abstract{Measurements are presented of the \PW and \PZ boson production cross sections in proton-proton collisions at a center-of-mass energy of 13.6\TeV. Data collected in 2022 and corresponding to an integrated luminosity of 5.01\fbinv with one or two identified muons in the final state are analyzed. The results for the products of total inclusive cross sections and branching fractions for muonic decays of \PW and \PZ bosons are $11.93\pm0.08\syst\pm0.17\lum\,^{+0.07}_{-0.07}\acc\unit{nb}$ for \PWp boson production, $8.86\pm0.06\syst\pm0.12\lum\,^{+0.05}_{-0.06}\acc\unit{nb}$ for \PWm boson production, and $2.021\pm0.009\syst\pm0.028\lum\,^{+0.011}_{-0.013}\acc\unit{nb}$ for the \PZ boson production in the dimuon mass range of 60--120\GeV, all with negligible statistical uncertainties. Furthermore, the corresponding fiducial cross sections, as well as cross section ratios for both fiducial and total phase space, are provided. The ratios include charge-separated results for \PW boson production (\PWp and \PWm) and the sum of the two contributions (\PWpm), each relative to the measured \PZ boson production cross section. Additionally, the ratio of the measured cross sections for \PWp and \PWm boson production is reported. All measurements are in agreement with theoretical predictions, calculated at next-to-next-to-leading order accuracy in quantum chromodynamics.}

\hypersetup{%
pdfauthor={CMS Collaboration},%
pdftitle={Measurement of the inclusive W and Z boson production cross sections and their ratios in proton-proton collisions at sqrt(s) = 13.6 TeV},%
pdfsubject={CMS},%
pdfkeywords={CMS, standard model, W and Z boson, cross section}}

\maketitle

\section{Introduction}

The massive \PW and \PZ bosons are fundamental components of the electroweak (EWK) sector of the standard model (SM) of elementary particle physics.
The precise determination of their production cross sections in proton-proton (\pp) collisions at the CERN LHC is important for probing theoretical predictions, particularly in high orders of perturbative quantum chromodynamics (QCD).
Measurements of these cross sections and other parameters of the SM are important in the search for unknown physics beyond the SM, where the productions of \PW and \PZ bosons are background processes.
This is also relevant when exploring new energy regimes by increasing the center-of-mass energy of the collider.

This article describes measurements of the inclusive total and fiducial \PW and \PZ boson production cross sections, as well as their ratios, at $\sqrt{s}=13.6\TeV$. Previous analyses by the ATLAS, CMS, and LHCb Collaborations have reported results at $\sqrt{s}=2.76$, 5.02, 7, 8, 13, and 13.6\TeV~\cite{Aad:2019bdc,CMS:2014dyj,Aaboud:2018nic,CMS:2024myi,Aaboud:2016btc,Khachatryan:2010xn,CMS:2011aa,Aaij:2014wba,Aaij:2012mda,Aaij:2015gna,Aaij:2016qqz,Aaij:2016mgv,Aaij:2015zlq,Chatrchyan:2014mua,Aad:2016naf,ATLAS:2024irg}.
The size of the analyzed data set corresponds to an integrated luminosity of $5.01\pm0.07\fbinv$, collected when the LHC began to collide protons at the unprecedented center-of-mass energy of 13.6\TeV in 2022.
Despite the statistical limitations introduced by the focus on muonic final states, and the relatively small size of this data set collected, the precision of the cross section results exceeds that of previous measurements, which can be attributed to a more precise determination of the integrated luminosity.

Events with one or two identified muons in the final state are analyzed.
The \DY Drell--Yan process is measured in the $\DY\to\MM$ channel for events with an invariant mass of the dimuon pair in the range 60--120\GeV.
Since the dominating process in this mass range comes from the production of \PZ bosons, the \DY process is referred to as \PZ boson production in the following.
The cross sections for the production of \PW and \PZ bosons are measured in fiducial regions, matching the experimentally accessible phase space, and extrapolated to the total cross sections in the full phase space by applying acceptance corrections.
In addition, ratios of cross sections are determined with higher precision than the individual cross section results, as several systematic uncertainties---most notably the uncertainty in the integrated luminosity measurement---cancel out when measuring the ratios.
Numerical values from all figures are provided in the HEPData record for this analysis~\cite{hepdata}.

\section{The CMS detector}

The central feature of the CMS apparatus is a superconducting solenoid of 6\unit{m} internal diameter, providing a magnetic field of 3.8\unit{T}.
Within the solenoid volume are a silicon pixel and strip tracker, a lead tungstate crystal electromagnetic calorimeter (ECAL), and a brass and scintillator hadron calorimeter (HCAL), each composed of a barrel and two endcap sections.
Forward calorimeters extend the pseudorapidity coverage provided by the barrel and endcap detectors.
Muons are detected in gas-ionization detectors embedded in the steel flux-return yoke outside the solenoid, and measured in both the tracker and the muon system.
A more detailed description of the CMS detector, together with a definition of the coordinate system used and the relevant kinematic variables, is reported in Refs.~\cite{CMS:2008xjf,CMS:2023gfb}.

Events of interest are selected using a two-tiered trigger system.
The first level (L1), composed of custom hardware processors, uses information from the calorimeters and muon detectors to select events at a rate of around 100\unit{kHz} within a fixed latency of about 4\mus~\cite{CMS:2020cmk}.
The second level, known as the high-level trigger (HLT), consists of a farm of processors running a version of the full event reconstruction software optimized for fast processing, and reduces the event rate to around 5\unit{kHz} before data storage~\cite{CMS:2016ngn,CMS:2024aqx}.

The particle-flow (PF) algorithm~\cite{CMS:2017yfk} reconstructs and identifies each particle in an event, with an optimized combination of information from the various elements of the CMS detector.
The reconstructed vertex with the largest value of summed transverse momentum squared $\pt^2$ of physics objects is defined as the primary \pp interaction vertex (PV)~\cite{CMS-TDR-15-02}.
The energy of photons is obtained from the ECAL measurement.
The energy of electrons is determined from a combination of the electron momentum at the PV as determined by the tracker, the energy of the corresponding ECAL cluster, and the energy sum of all bremsstrahlung photons spatially compatible with originating from the electron track.
The energy of muons is obtained from the curvature of the corresponding track.
The energy of charged hadrons is determined from a combination of their momentum measured in the tracker and the matching ECAL and HCAL energy deposits, corrected for the response function of the calorimeters to hadronic showers.
Finally, the energy of neutral hadrons is obtained from the corresponding corrected ECAL and HCAL energies.

The missing transverse momentum vector \ptvecmiss is computed as the negative vector sum of the transverse momenta of all the PF candidates in an event from the PV, and its magnitude is denoted as \ptmiss~\cite{CMS:2019ctu}.
The \ptvecmiss is modified to account for corrections to the energy scale of the reconstructed jets in the event.

\section{Data and Monte Carlo simulation}

This analysis uses data collected with the CMS detector during part of the 2022 LHC run at a 13.6\TeV center-of-mass energy.
Events are selected based on the presence, at the HLT level, of at least one isolated muon with a transverse momentum $\pt>24\GeV$ and a pseudorapidity $\abseta<2.4$.

The signal samples for modeling the \PW and \PZ boson production are generated in a Monte Carlo (MC) simulation using \MGvATNLO version 2.9.9~\cite{Alwall:2014hca} at next-to-leading order (NLO) in perturbative QCD with up to two outgoing partons at Born level, where the different jet multiplicities have been merged with the FxFx technique~\cite{Frederix:2012ps}.
Background processes with top quarks, \ie, the production of top quark-antiquark pairs (\ttbar) and the production of single top quarks via the $t$ channel or in association with a \PW boson, as well as diboson events ($\PW\PW$, $\PW\PZ$, $\PZ\PZ$, collectively referred to as \VV) are simulated at NLO using \POWHEG version 2~\cite{Nason:2004rx,Frixione:2007vw,Alioli:2010xd,Frederix:2012dh,Re:2010bp,Nason:2013ydw}.
For single top quark events in the $t$ channel, \textsc{MadSpin}~\cite{Artoisenet:2012st} is used to describe the top quark decays.
Events with electroweak production of a \PZ or \PW boson and two jets are simulated with \MGvATNLO version 2.9.13 at leading-order (LO) precision in perturbation theory.
The \PYTHIA 8.306 package~\cite{Bierlich:2022pfr} is used for the simulation of parton showering, hadronization, and the underlying event, using tune CP5~\cite{CMS:2019csb}.
In all generated events, the proton structure is described by the parton distribution functions (PDFs) at next-to-NLO (NNLO) provided by NNPDF 3.1~\cite{NNPDF:2017mvq}.
For all processes, the detector response is simulated with the \GEANTfour toolkit~\cite{Agostinelli:2002hh}.
Additional \pp interactions within the same or nearby bunch crossings (pileup) are simulated with minimum bias and included in all MC simulation samples.

\section{Event selection}

The decay of the \PW boson in the muon channel is characterized by the presence of a prompt, energetic, and isolated muon, and large \ptmiss originating from the undetected neutrino.
The muonic \PZ boson decay features two prompt, energetic and isolated muons with opposite electric charges, with an invariant mass of the two muons close to the mass of the \PZ boson, and no intrinsic \ptmiss in the event.
In the analysis, only decays into muons are considered, as they can be identified more efficiently and their \pt can be measured more precisely by the CMS detector compared with electrons or tau leptons~\cite{CMS:2018rym,CMS:2020uim,CMS-DP-2020-021}. Electrons and taus still indirectly contribute to selected events in the calculation of the missing transverse momentum. The impact from electrons in jets and hardonically decaying taus on the missing transverse momentum is mitigated through a dedicated hadronic recoil correction explained in Sec.~\ref{sec:calibration}. The excellent muon performance has also been demonstrated in the recent measurement of the \PW boson mass by the CMS experiment~\cite{CMS:2024lrd}.
Reconstructed muon candidates are required to satisfy a set of quality criteria corresponding to the ``tight'' working point of the muon identification~\cite{CMS:2018rym}.
The criteria are based on the number of spatial measurements in the silicon tracker and the muon system, the fit quality of the combined muon track, as well as the transverse impact parameter.
These criteria largely suppress nonprompt muons from hadron decays and hadron shower remnants in the muon chambers.
In addition, muon candidates must be isolated from other activity in the event. The relative muon isolation is defined as the scalar sum of the \pt of all particles reconstructed by the PF algorithm, except for the muon, within a cone of size $\DR=0.4$ around the muon track, with $\DR=\sqrt{\smash[b]{(\Delta\eta)^2+(\Delta\phi)^2}}$, and divided by the muon \pt.
The relative isolation of the muon, corrected for pileup contributions, must be smaller than 0.15 for this analysis.
In addition, the reconstructed muon candidates must satisfy $\pt>25\GeV$ and $\abseta<2.4$, ensuring high efficiency of the muon trigger used to collect the data sample.

The \PW boson candidate events are required to have exactly one identified muon satisfying the criteria described above.
Events with a second muon, identified at the ``loose'' working point, are rejected, reducing the contamination from \PZ boson events.
Muons passing the loose working point are only required to be reconstructed with the PF algorithm and to be matched to a track in the tracker detector~\cite{CMS:2018rym}.
Although \PW boson events feature large missing transverse momentum, no requirement on that is made at this stage.
Instead, this information is used later on for the signal extraction.
The transverse mass \mT, defined as
\begin{equation}
\mT = \sqrt{2\pt^{\PGm}\ptmiss(1-\cos(\Delta\phi))},
\end{equation}
with the muon transverse momentum $\pt^{\PGm}$, and $\Delta\phi$ being the azimuthal opening angle between the muon and the \ptvecmiss direction, is a sensitive kinematic variable in this channel, used to distinguish signal from background contributions.

For the selection of \PZ boson candidate events, two muons satisfying the above described criteria with opposite electric charges are required. Events with additional muons satisfying these criteria are rejected in order to reduce contributions from diboson productions.
The invariant mass of the muon pair \mmumu is required to be near the \PZ boson mass, within the range 60--120\GeV.
This requirement reduces the relative contribution of nonresonant Drell--Yan production to the signal.

The aim of the analysis is the measurement of the inclusive cross sections of \PW and \PZ boson production, hence no requirements on the presence of accompanying hadronic jets in the event are made.

The fiducial regions for the \PW and \PZ boson acceptances are defined in the simulated events by a set of kinematic requirements applied to the final states at the generator level, emulating the event selection criteria applied at the reconstruction level.
For the \PW boson production, the fiducial region is defined by the aforementioned kinematic \pt and \sigeta requirements on the muon, applied before the emission of final-state radiation (FSR).
As opposed to the theoretical definition, the event selection in the fiducial measurements is applied at reconstruction level, where the muons may have undergone FSR effects. The effect on the acceptance is at the level of 2\% for \PW boson production, and at the level of 4\% for \PZ boson production and is corrected to ensure consistency in the comparison to theoretical predictions.

\section{Analysis strategy}\label{sec:strategy}

The product of the fiducial cross section for \PW and \PZ boson production and the muonic branching fraction can be expressed in a generalized way as:
\begin{equation}
\sigfid \BR  = \frac{N_{\text{sig}}}{\epsilon L} = \frac{N - N_{\text{bkg}}}{\epsilon L},
\end{equation}
where \BR is the branching fraction for either \WptoMpN, \WmtoMmN, or \ZtoMM, $N$ is the number of observed events, $N_{\text{sig}}$ and $N_{\text{bkg}}$ are the number of signal and background events, respectively, $\epsilon$ is the efficiency of the signal events within the fiducial region satisfying the event selection requirements, and $L$ is the integrated luminosity.

In total, three signal regions are used for the analysis: two single-muon regions (\PGmp and \PGmm) and one double-muon region (\MM).
In the single-muon regions, the transverse mass \mT is used to distinguish the \PWpm boson signal from background processes, and \mmumu is used for that purpose in the double-muon signal region for the \PZ boson signal.
All three distributions are fitted simultaneously by means of a binned maximum likelihood fit.

The results for the product of the cross sections and branching fractions in the full phase space are then calculated by dividing the expression for the fiducial region, which is already aligned with the theoretical definition of the signal, by the kinematic acceptance $A$ of this region:
\begin{equation}
\sigtot\BR = \frac{\sigfid\BR}{A}.
\end{equation}
The kinematic acceptance $A$ is derived by calculating the ratio of the theoretically predicted signal cross sections with and without the selection requirements corresponding to the definition of the fiducial region:
\begin{equation}
  A = \frac{(\sigfid\BR)_{\text{predicted}}}{(\sigtot\BR)_{\text{predicted}}}.
\end{equation}
For the acceptance in the \PZ boson region, the total cross section of \PZ boson production in the mass range 60--120\GeV is calculated without any further requirements on the muons and enters the definition of the acceptance at the denominator.
For the calculation of the fiducial cross section, which enters the acceptance at the numerator, the muons must additionally fulfill $\pt>25\GeV$ and $\abseta<2.4$, emulating the event selection criteria at the generator level.
For the acceptance calculation in the \PW boson regions, the procedure is repeated analogously with a phase space restriction on the mass of the \PW boson of $m_{\PW}>20\GeV$, which ensures the convergence of the resummation calculations without leading to a visible change in the cross section values.
The cross sections are computed numerically at NNLO QCD and next-to-next-to-leading logarithmic (NNLL) \qt resummation, where \qt is the transverse momentum of the \PW or \PZ boson, with the program \DYTurbo 1.3.2~\cite{Camarda:2019zyx,Camarda:2021ict,Camarda:2021jsw,Camarda:2023dqn} and the NNPDF 3.1~\cite{NNPDF:2017mvq} PDF sets.
The following ratios are provided as well:
\begin{align}
  R^{\WpoverZ} &= \frac{\sigma(\pp\to\PWp + \PX)\BR(\WptoMpN)}{\sigma(\pp\to\PZ + \PX)\BR(\ZtoMM)}, \\
  R^{\WmoverZ} &= \frac{\sigma(\pp\to\PWm + \PX)\BR(\WmtoMmN)}{\sigma(\pp\to\PZ + \PX)\BR(\ZtoMM)}, \\
  R^{\WpmoverZ} &= \frac{\sigma(\pp\to\PWpm + \PX)\BR(\PWpm\to\MpN/\MmN)}{\sigma(\pp\to\PZ + \PX)\BR(\ZtoMM)}, \\
  R^{\WpoverWm} &= \frac{\sigma(\pp\to\PWp + \PX)\BR(\WptoMpN)}{\sigma(\pp\to\PWm + \PX)\BR(\WmtoMmN)},
\end{align}
where some of the systematic uncertainties, including the uncertainty in the integrated luminosity, cancel out.
The NNLO+NNLL predictions for the fiducial and total inclusive cross sections are shown in Table~\ref{tab:theoryxsec}, the corresponding cross section ratios in Table~\ref{tab:theoryratio}.
The PDF uncertainties in the tables are calculated with the NNPDF 3.1 PDF sets. The scale uncertainties, representing missing higher-order contributions in QCD, are evaluated by varying the renormalization and factorization scales. These scales are set to the mass of the dimuon system as the nominal value and varied up and down by a factor of two (omitting the extreme case where both scales are varied in opposite directions). The final scale uncertainty is taken from the maximum deviation. For the cross section ratios, the different scale variations are treated as fully correlated between the cross sections, leading to partial cancellations. The integration uncertainties reflect the uncertainties of the numerical calculation of integrals in \DYTurbo. The asymmetric behavior of the scale uncertainties is caused by outliers from different variations of the individual renormalization and factorization scales.

\begin{table}[!ht]
\centering
\topcaption{Predictions for the product of the fiducial and total inclusive cross sections and branching fractions. The first uncertainty is the PDF uncertainty, the second is the QCD scale uncertainty of the calculation, and the third is the integration uncertainty.}
\label{tab:theoryxsec}
\renewcommand{\arraystretch}{1.3}
\begin{tabular}{lcl@{\,$\pm$\,}l@{\,}l@{\,$\pm$\,}lccl@{\,$\pm$\,}l@{\,}l@{\,$\pm$\,}lc}
    \hline
    Process & \multicolumn{6}{c}{$\sigfid\BR\pm\PDFns\pm\scalens\pm\intns$ in nb} & \multicolumn{6}{c}{$\sigtot\BR\pm\PDFns\pm\scalens\pm\intns$ in nb} \\
    \hline
    \PWp  && 5.490 & 0.040 & ${}^{+0.025}_{-0.048}$ & 0.007 &&& 12.06 & 0.10 & ${}^{+0.06}_{-0.13}$ & 0.01 \\
    \PWm  && 4.222 & 0.032 & ${}^{+0.014}_{-0.045}$ & 0.005 &&& \phantom{0}8.97 & 0.07 & ${}^{+0.05}_{-0.10}$ & 0.01 \\
    \PWpm && 9.710 & 0.071 & ${}^{+0.033}_{-0.093}$ & 0.009 &&& 21.03 & 0.16 & ${}^{+0.11}_{-0.23}$ & 0.01 \\
    \PZ   && 0.7663 & 0.0067 & ${}^{+0.0031}_{-0.0050}$ & 0.0007 &&& 2.027 & 0.015 & ${}^{+0.008}_{-0.019}$ & 0.001 \\
    \hline
\end{tabular}
\end{table}

\begin{table}[!ht]
\centering
\topcaption{Predictions for the fiducial and total inclusive cross section ratios. The first uncertainty is the PDF uncertainty, the second is the QCD scale uncertainty of the calculation, and the third is the integration uncertainty.}
\label{tab:theoryratio}
\renewcommand{\arraystretch}{1.3}
\begin{tabular}{lr@{\,$\pm$\,}l@{\,}l@{\,$\pm$\,}lr@{\,$\pm$\,}l@{\,}l@{\,$\pm$\,}l}
    \hline
    Process & \multicolumn{4}{c}{$\Rfid\pm\PDFns\pm\scalens\pm\intns$} & \multicolumn{4}{c}{$\Rtot\pm\PDFns\pm\scalens\pm\intns$} \\
    \hline
    \WpoverZ & 7.158 & 0.041 & ${}^{+0.031}_{-0.016}$ & 0.012 & 5.951 & 0.027 & ${}^{+0.008}_{-0.010}$ & 0.004 \\
    \WmoverZ & 5.508 & 0.030 & ${}^{+0.003}_{-0.025}$ & 0.009 & 4.424 & 0.016 & ${}^{+0.005}_{-0.006}$ & 0.004 \\
    \WpmoverZ & 12.666 & 0.069 & ${}^{+0.034}_{-0.038}$ & 0.016 & 10.375 & 0.040 & ${}^{+0.010}_{-0.016}$ & 0.007 \\
    \WpoverWm & 1.300 & 0.003 & ${}^{+0.005}_{-0.000}$ & 0.002 & 1.345 & 0.004 & ${}^{+0.003}_{-0.001}$ & 0.001 \\
    \hline
\end{tabular}
\end{table}

\section{Simulation tuning and data calibration}\label{sec:calibration}

The conditions used for the simulation of the detector are not necessarily identical to the actual conditions during data taking.
To mitigate the bias from these differences, several corrections are applied to the simulation.
Furthermore, a muon momentum scale calibration as well as a \ptmiss correction are applied to the data.

Differences in the pileup between simulation and data may bias the result of the analysis.
To adjust the pileup of simulated events to match the actual conditions during \pp collisions, dedicated correction factors are derived from comparing the simulated pileup distribution with the luminosity-based pileup distribution in the detected events.
The latter is calculated from the measured instantaneous luminosity in fixed intervals during data taking, and the measured value of the inelastic \pp cross section.
A weight is calculated to correct the simulated pileup distribution to the one in data.
After reweighting the simulated events, the agreement between simulation and data in the distribution of the number of well-reconstructed primary vertices is improved and well within the uncertainties mentioned in Section~\ref{sec:systematics}.

To account for higher-order effects not present in the MC generators, the simulation for \PZ boson production is reweighted based on corrections obtained from the difference in the \pt distribution of the dimuon system between data and simulation.
The same corrections are applied to the simulation of \PW boson production.

Since the accurate determination of the muon identification, isolation, and reconstruction efficiencies is a crucial part of the cross section measurements, scale factors compensating for the differences between simulation and data are derived for each selection step in bins of \sigeta and \pt of the muon.
A tag-and-probe method~\cite{Khachatryan:2010xn} is employed for this purpose on the dimuon sample in both data and simulation, with backgrounds subtracted from the data to obtain a pure sample.
Tag muons are defined as muons passing the analysis selection requirements as well as being matched to the appropriate trigger object.
The exact definition of probe muons depends on the efficiency under study.
They are selected among the muons fulfilling the \pt and \sigeta requirements of the event selection.
The efficiency in the tag-and-probe method is defined as the number of probe muons passing a selection criterion divided by the sum of all probe muons before the corresponding selection.

Assuming that the muon efficiency can be factorized as
\begin{equation}
\efficiency{\PGm} = \efficiency{Trk}\,\efficiency{Sta}\,\efficiency{ID}\,\efficiency{Iso}\,\efficiency{Trigger},
\end{equation}
the efficiencies are studied one after the other and the definition of the probe muon is given by the passing probe definition of the preceding efficiency stage.
The efficiencies under study are the efficiency of the track reconstruction in the inner tracker \efficiency{Trk}, the reconstruction efficiency in the muon detectors (standalone) \efficiency{Sta}, the muon identification efficiency \efficiency{ID}, the muon isolation efficiency \efficiency{Iso}, and the efficiency of the muon trigger \efficiency{Trigger}.
For both data and simulation, the efficiencies are determined in bins of \pt and \sigeta of the muon and separately for the two electric charges in the case of the trigger efficiency.
The charge-dependent difference in trigger efficiency for the collected events is at the per-mille level in the central part of the detector, but increases to the low percent level in the forward region.
The scale factors are then obtained as the ratio of the efficiency values in data and simulation and are applied as event weights to the simulated events, depending on the \pt, \sigeta and the sign of the identified muons in the event.

Further differences between data and simulation arise from the potential occurrence of false assignments of detector information to the preceding bunch crossing~\cite{CMS:2021yvr}.
In such cases, the trigger accepts the event in the bunch crossing preceding the actual bunch crossing of interest, and the data of the interesting event are not recorded.
To account for this effect, the corresponding muon trigger inefficiency has been measured in data and is applied as a scale factor to simulation.
In the central area of the detector, the inefficiency is at the per-mille level, increasing to the percent level in the outer regions.

Misalignments and miscalibrations of the detector, uncertainties in the magnetic field of the detector, as well as shortcomings in the simulation of the relevant processes can cause discrepancies in lepton momentum scale and resolution between data and simulation.
Muon momentum scale and resolution corrections are derived by comparing the \PZ boson peak position in the distributions of the invariant mass of the dimuon pair between data and simulation.
Correction factors for momentum scale and resolution are derived by fitting the \mmumu distributions in multiple bins of muon \pt and \sigeta.
The signal shape is parameterized by convolving a Breit--Wigner function and a double-sided Crystal Ball function~\cite{oreglia1980,gaiser1982}.
The mean and the width of the Breit--Wigner function are fixed to the known mass and width of the \PZ boson.
The Gaussian part of the double-sided Crystal Ball function is symmetric and models detector effects, while its exponential sides account for further effects, such as radiative tails.

The momentum scale correction factors are given by the ratio of the fitted peak positions of the \mmumu distributions in simulation and data, $\langle\mmumu^{\text{MC}}\rangle$ and $\langle\mmumu^{\text{data}}\rangle$.
These factors, typically at the per-mille level, are applied to the \pt of muons in data:
\begin{equation}
\pt^{\text{data,\ corr.}}=\pt^{\text{data\vphantom{,}}}\frac{\langle\mmumu^{\text{MC}}\rangle}{\langle\mmumu^{\text{data}}\rangle}.
\end{equation}
The total width of the fitted peak in the \mmumu distribution consists of the intrinsic decay width of the \PZ boson and the additional contribution from the detector resolution, which is smaller in the simulation compared with data.
To mitigate this effect, an additional Gaussian smearing is introduced and applied to the simulated samples to match the momentum resolution observed in data.
The width of the Gaussian smearing varies, ranging from zero (indicating no smearing is required) up to 1.5\%, with an average value of around 0.6\%.

In addition, a precise reconstruction of the muon momentum, the accurate reconstruction of the missing transverse momentum of an event is essential for this analysis, as \ptmiss is used through \mT to distinguish \PW boson production from the background from SM events composed uniquely of jets produced through the strong interaction (QCD multijet events).
The predictions for \ptmiss rely on the modeling of the \PW boson recoil and detector effects, which can be difficult to simulate accurately.
Possible sources of discrepancy between simulation and data could result from deficiencies in the modeling of the calorimeter response and resolution or from incomplete description of the underlying event, as well as the displacement of the interaction point.
An observed modulation of \ptmiss in the azimuthal angle is mitigated using corrections based on the average value of \ptmiss in the $x$ and $y$ direction.
This correction is applied to both data and simulation.
Further residual discrepancies between data and simulation are addressed through corrections derived from a study of the \PZ boson recoil.
The recoil is decomposed into two components relative to the \PZ boson momentum: one parallel and one perpendicular to the boson direction.
Corrections are then obtained by aligning the cumulative distribution functions of these recoil components between data and simulation in bins of the \PW and \PZ boson momentum.

\begin{figure}[!ht]
    \centering
    \includegraphics[width=0.49\textwidth]{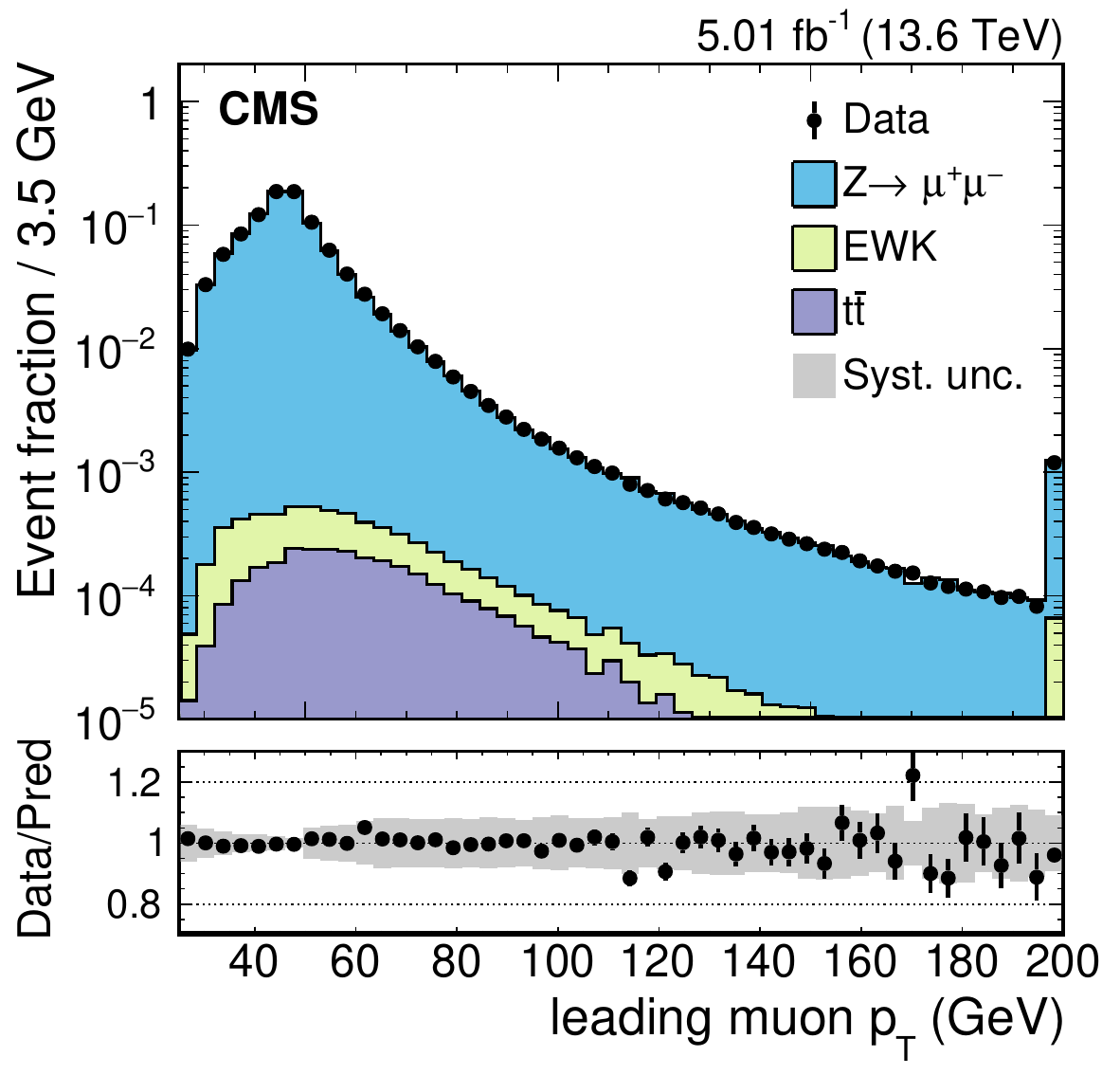}
    \includegraphics[width=0.49\textwidth]{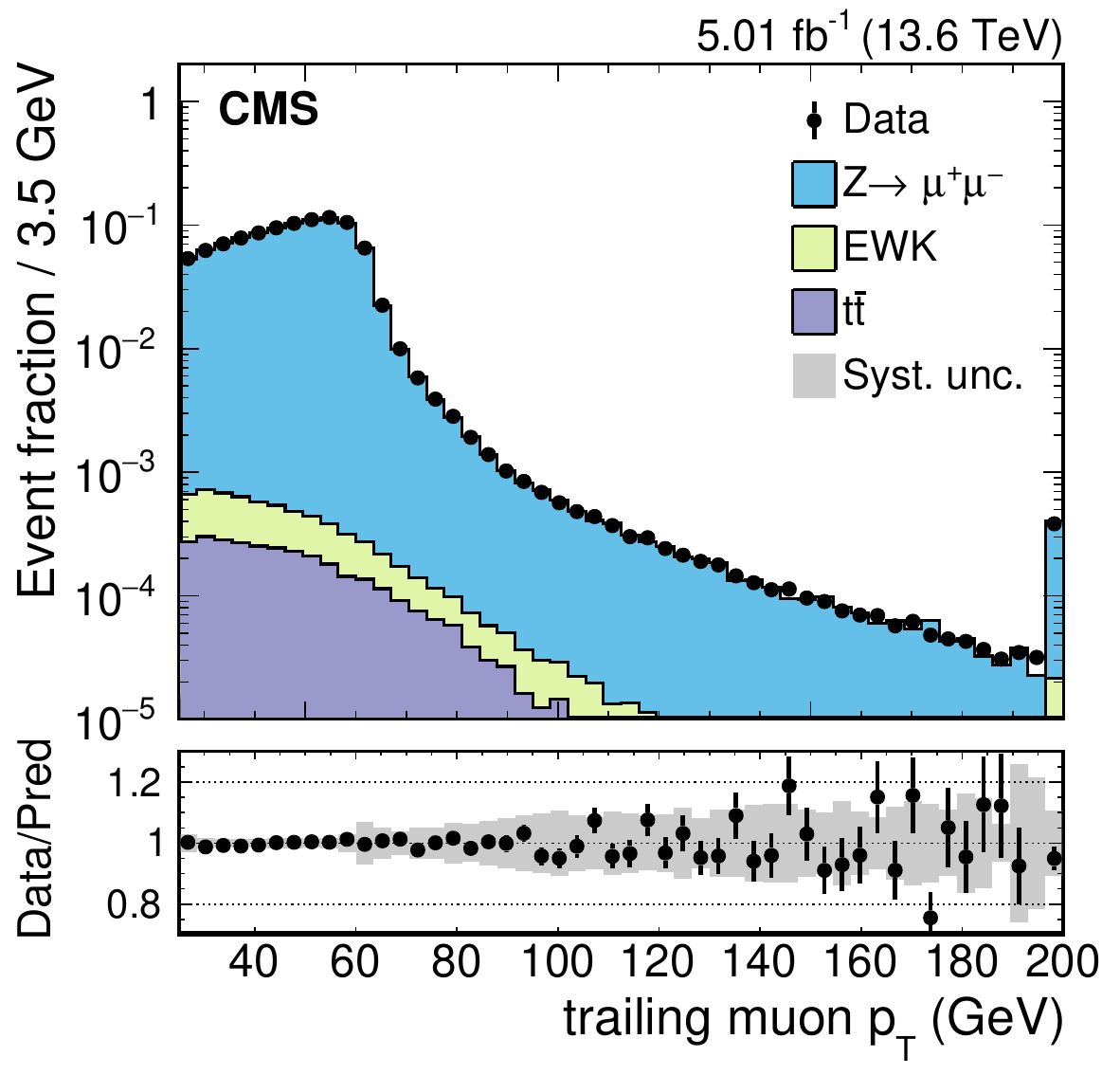}
    \includegraphics[width=0.49\textwidth]{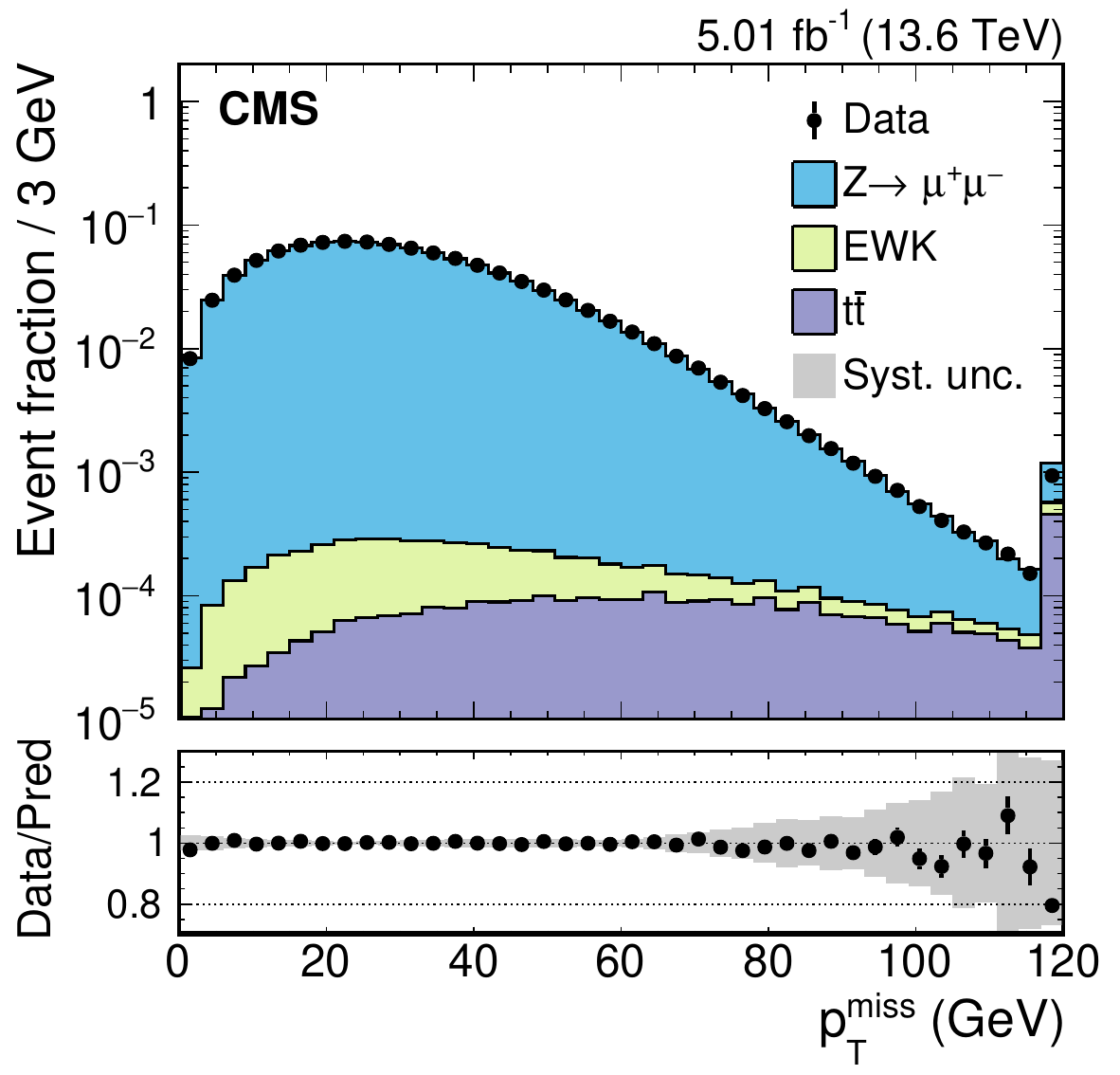}
    \caption{Corrected distributions of the \pt of the leading (upper left) and trailing (upper right) muon, as well as \ptmiss (lower) in the \PZ boson signal region. The distributions of both data and combined simulation in the upper panels are normalized because the scale of the signal sample is extracted later in the fit. The lower panels show the ratio of the normalized data to the normalized prediction. Overflow entries are included in the rightmost bin, respectively. Differences in the shape between data and prediction are fully covered by the systematic uncertainty band. The systematic uncertainties are described in detail in Section~\ref{sec:systematics}. The vertical uncertainty bars on the data represent the statistical uncertainty.}
    \label{fig:control_plots}
\end{figure}

The fully corrected distributions of the leading (highest \pt) and trailing (second highest \pt) muon transverse momenta, as well as the \ptmiss, both in the \PZ boson signal region, are shown in Fig.~\ref{fig:control_plots}.

\section{Background modeling}

Various background processes must be considered in the analysis, which are grouped into three distinct categories: \ttbar, EWK, and QCD multijet production.
Background categories with prompt muons relevant for this analysis are divided into  \ttbar and EWK categories, whereas nonprompt muons contribute to the QCD category.

The \ttbar category consists of the production of top-quark-antiquark pairs.
The production of single top quarks, \VV production, the EWK production of a \PW or \PZ boson (\PV) via vector boson fusion in association with two jets (VBF \PV), and contributions from \PW or \PZ bosons decaying into \PGt leptons, which subsequently decay to muons, are considered in the EWK category.
All these background processes are modeled using samples of simulated events.
The contribution from \PW bosons plus hadrons misidentified as a muon in the (\MM) signal region is negligible.

{\tolerance=800
The third background category, QCD multijet events, comes from SM events composed uniquely of jets produced from the strong interaction.
The muon comes from a weak decay within one of the jets or from a jet misidentified as a muon when the jet is not fully absorbed in the HCAL and reaches the muon detectors.
This background contribution is only present in the \PW boson selection, where it is already greatly reduced by the isolation requirement on the muon candidates.
However, because of the large production cross section, the remaining QCD multijet events represent the largest background contribution to the single-muon signal regions.
Although the normalization of the QCD multijet background is determined in the fit, its shape as a function of \mT is modeled using data events from a QCD-enriched control region.
This region is defined by the same event selection requirements as the \PW boson signal region, with the exception of the muon isolation requirement, which is inverted.
\par}

\begin{figure}[!b]
    \centering
    \includegraphics[width=0.49\textwidth]{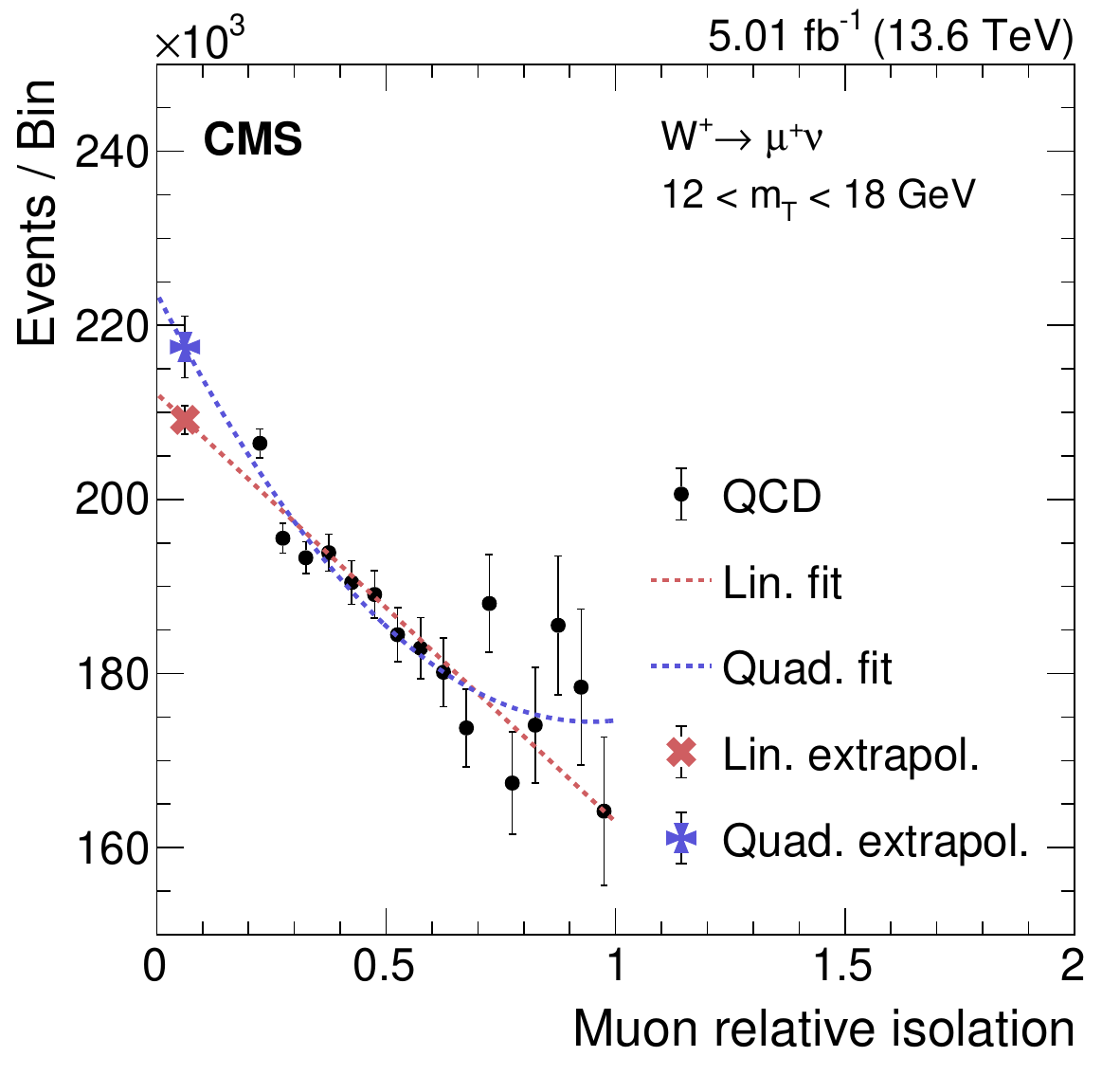}
    \includegraphics[width=0.49\textwidth]{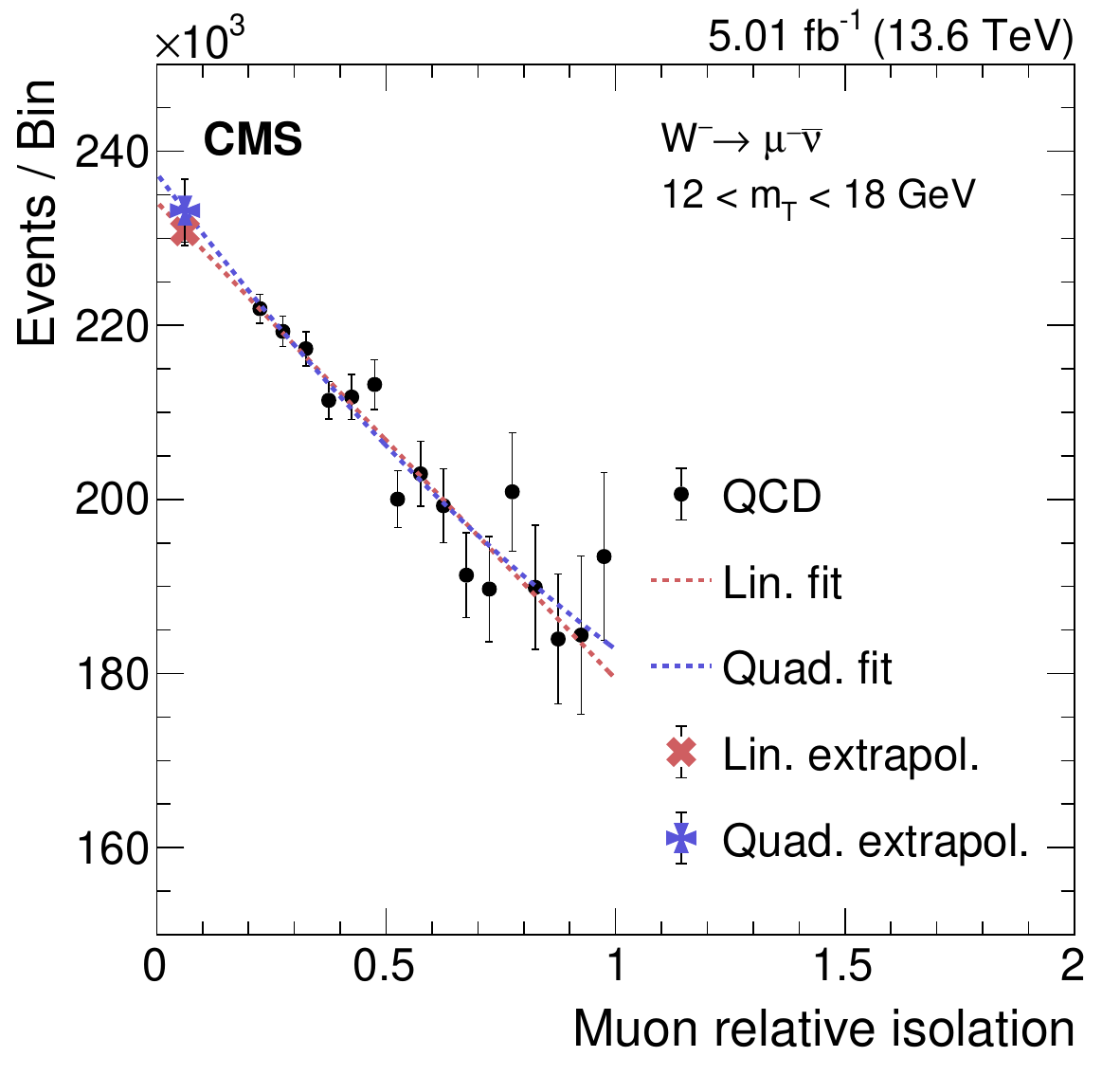}
    \includegraphics[width=0.49\textwidth]{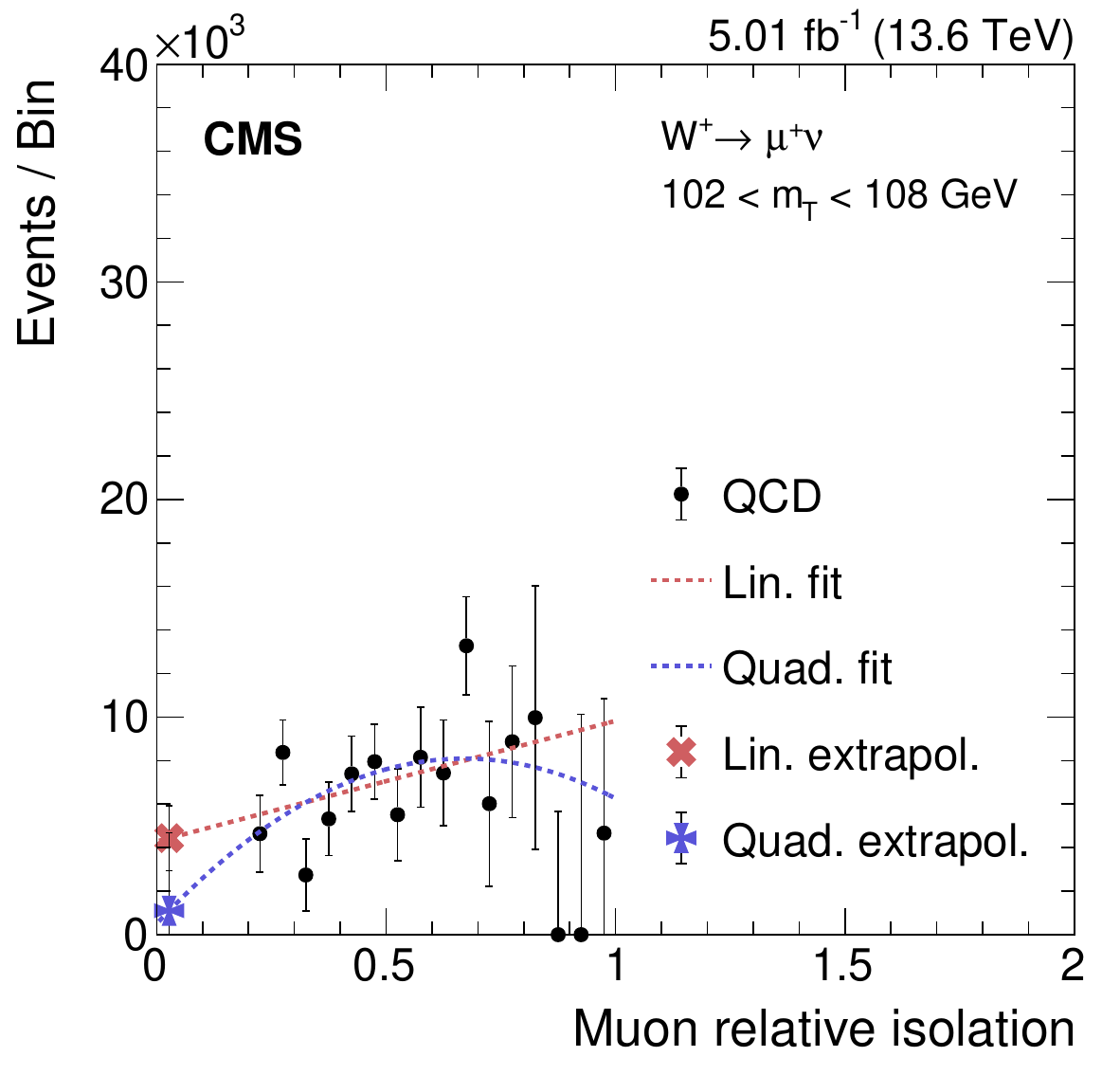}
    \includegraphics[width=0.49\textwidth]{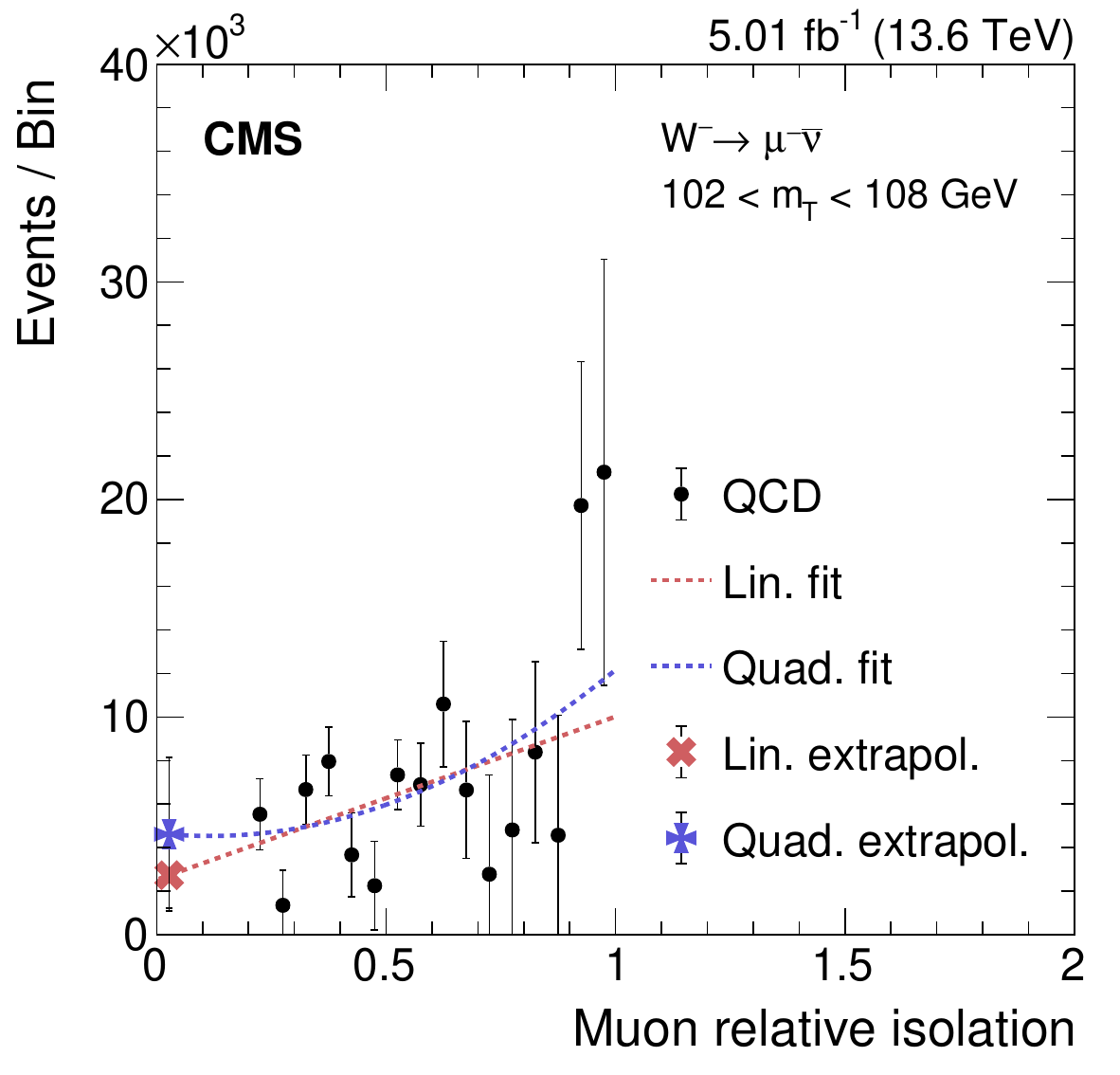}
    \caption{Example figures to illustrate the extrapolation procedure to estimate the QCD multijet background from the QCD-enriched control region. The yield in each bin is obtained by subtracting the simulated non-QCD contributions from the measured data. At high values of the transverse mass, the statistical uncertainty of the QCD yield increases, represented by the vertical uncertainty bars, leaving more freedom for the polynomial fit.}
    \label{fig:qcd_extrapolation}
\end{figure}

\begin{figure}[!ht]
    \centering
    \includegraphics[width=0.49\textwidth]{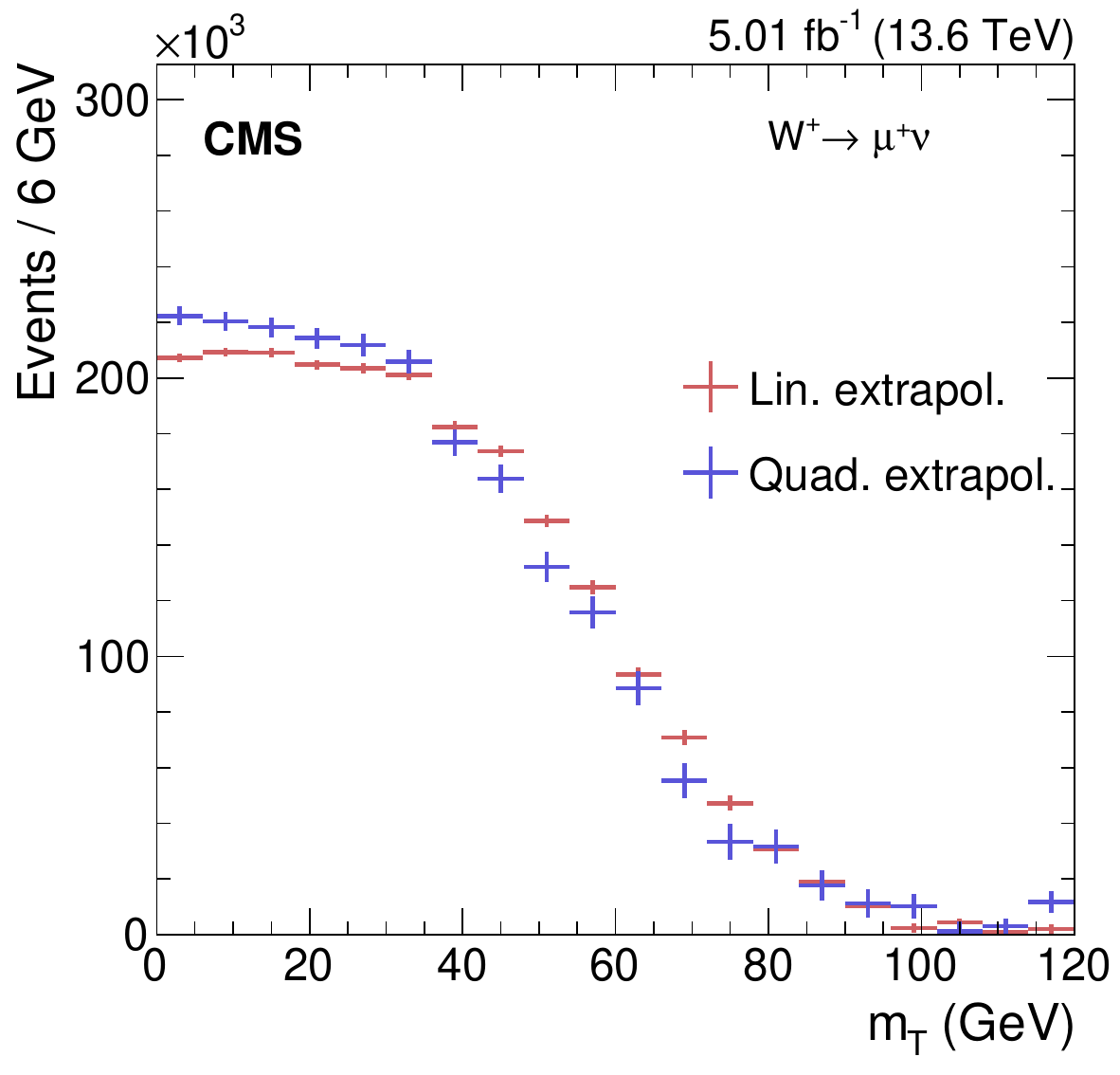}
    \includegraphics[width=0.49\textwidth]{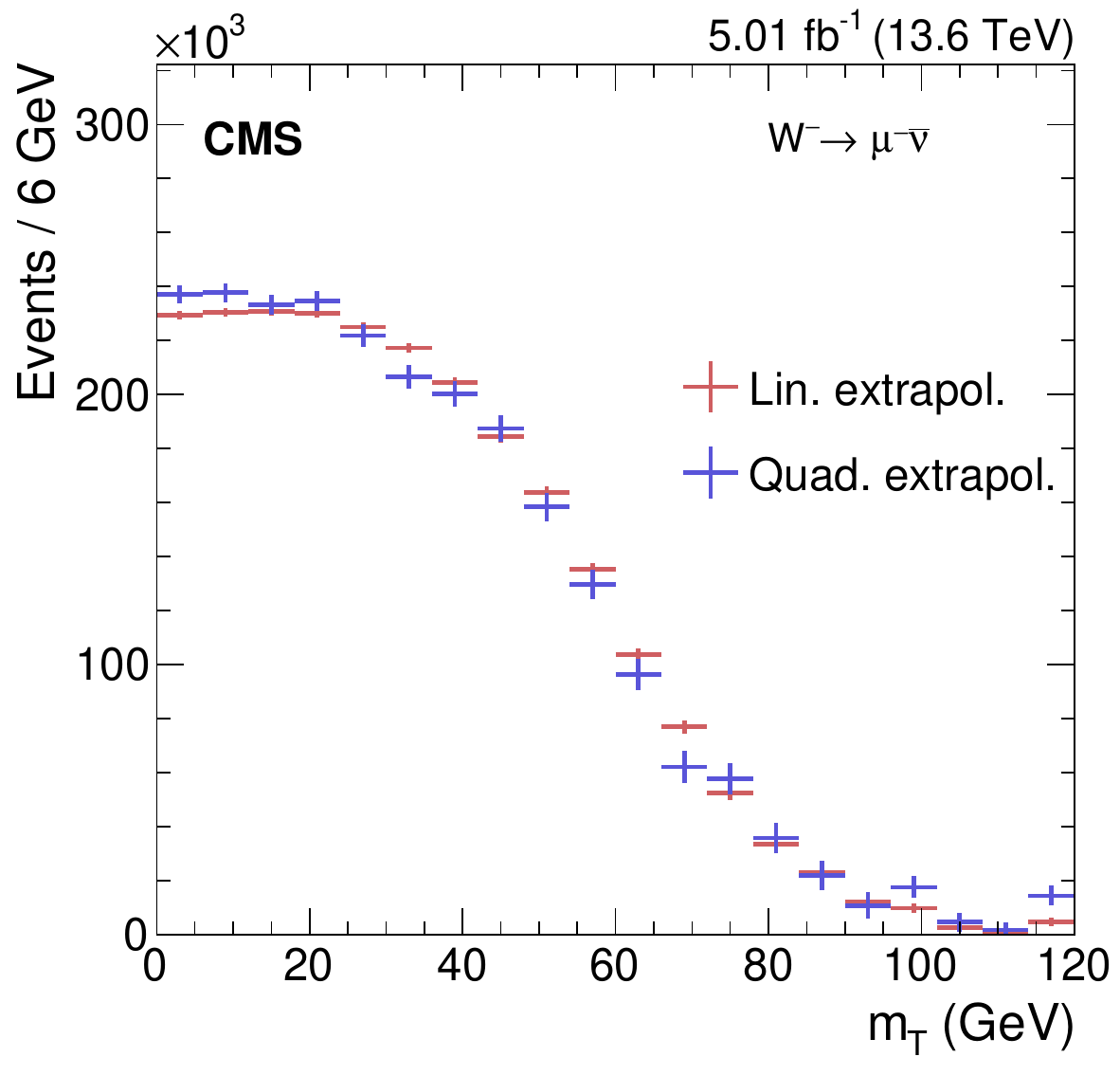}
    \caption{Result of the extrapolated \mT distribution for the QCD multijet background in the signal region from the QCD-enriched control region. The template normalization is left freely floating in the fit, as the final normalization is determined in the fit. The vertical uncertainty bars correspond to the extrapolation uncertainty from the fit.}
    \label{fig:qcd_templates}
\end{figure}

The remaining non-QCD contribution in this QCD-enriched control region is subtracted using simulated samples.
The QCD-enriched region is divided into several subsamples, depending on the value of the relative isolation of the muon in the event.
In total, 16 subsamples are created in the relative isolation range 0.20--1.00 with a step size of 0.05.
The \mT distributions for these subsamples are normalized to the same area, and for each \mT bin, the 16 corresponding values from the different subsamples are fitted to account for the dependence of the \mT distribution shape on the relative isolation of the muon observed in data.
The fit is performed using two different fit functions, a linear and a quadratic one.
Four fits in different \mT bins are exemplary shown in Fig.~\ref{fig:qcd_extrapolation}.
The extrapolation to the signal region (where the muon is isolated) is evaluated at the mean of the isolation distribution created by subtracting the distribution for non-QCD events from the distribution in data for each \mT bin.
The template derived from the linear fit function is chosen to be the nominal \mT template and the difference to the template derived from the quadratic fit function is symmetrized and incorporated as a contribution to the QCD template shape uncertainty.
Both templates are shown in Fig.~\ref{fig:qcd_templates}.
The assumption that the shape of the \mT distribution for QCD multijet events in the QCD-enriched control region is consistent with that in the single-muon signal regions is tested by splitting the control region.
Roughly half of the region, with muon relative isolation values between 0.5 and 1.0, is used to derive a QCD \mT template in the other half, with relative isolation values between 0.2 and 0.5.
The resulting template in the part of the control region with isolation values between 0.2 and 0.5 is then used to fit the \mT distribution.
The resulting post-fit distribution describes the data in the control region with a precision better than one percent.

\section{Signal extraction}

To extract the \PW and \PZ boson signal and background yields, a binned maximum likelihood fit is performed simultaneously to the transverse mass \mT distributions in the single-muon signal regions and the dimuon invariant mass \mmumu distribution in the double-muon signal region. The results are determined using the CMS statistical analysis tool \textsc{combine}~\cite{CMS:2024onh}, which is based on the \textsc{RooFit}~\cite{Verkerke:2003ir} and \textsc{RooStats}~\cite{Moneta:2010pm} frameworks.

The \mT and \mmumu templates are fitted to the observed distributions as the sum of three signal contributions---the \PWp, \PWm, and \PZ signals---as well as three distinct background contri\-butions---the \ttbar background, the EWK background, and the background from QCD multijet production. The signal and non-QCD background processes are modeled using simulation, with the simulated samples normalized using the theoretically predicted cross sections.
The \PZ boson sample is normalized using the theoretical cross section calculated at NNLO in QCD and NNLL \qt resummation with the program \DYTurbo and the NNPDF 3.1 PDF sets.
The cross sections of top quark-antiquark pair production, \PW boson production, single top quark production in the tW channel, and diboson production cross sections are taken from Ref.~\cite{CMS:2023qyl}.
Single top quark production in the $t$ channel has been calculated at NNLO in QCD with the \MCFM version 10.1 program~\cite{Campbell:2020fhf}, and the cross section of the EWK production of a \PZ boson and two jets is calculated at LO using \MGvATNLO.
The pre-fit normalization of the data-driven QCD distributions is taken from the extrapolated templates.
The event rates serve as inputs to the maximum likelihood fit and the individual normalization factors of the two QCD templates are finally determined through the fit itself, where they are allowed to vary freely.
The kinematic ranges in which the fits are performed are 0--120\GeV for \mT and 60--120\GeV for \mmumu.
Systematic uncertainties are included in the maximum likelihood fit as nuisance parameters.
The various sources of systematic uncertainty are detailed in the next section, where the individual values correspond to the width of the Gaussian constraint term for the respective nuisance parameter~\cite{Conway:2011in}.

\section{Systematic uncertainties}\label{sec:systematics}

Several sources of systematic uncertainty influence the results of this measurement.
The uncertainty in the muon efficiency scale factors contains statistical and systematic components.
The systematic uncertainty component is related to the model dependence of the approach used to extract the numbers of passing and failing probe muons from data.
To estimate the impact of this uncertainty source, alternative models for the description of signal and background are used, the \pt requirement on the tag muon is altered, and the mass binning and the fit range are varied.
Systematic as well as statistical uncertainty components are also included for the trigger inefficiency.
The uncertainty in the muon momentum correction is conservatively considered with 50\% of the total correction value.
The recoil correction comes with statistical and systematic uncertainty components.
The latter is estimated by using alternative fitting models, omitting the background subtraction, and varying the rapidity binning for the determination of the recoil corrections.
A 10\% uncertainty in the normalization of the backgrounds from simulation is assumed.
The model of the QCD multijet process is derived from the experimental data, where the uncertainties in the contamination from non-QCD processes in the QCD-enriched control region and in the choice of extrapolation function, as well as the statistical component, are included.
A potential bias in the method used to reweight the pileup distribution in the simulation is evaluated by varying the value of the inelastic \pp cross section.
The systematic uncertainty of the boson \pt reweighting is considered by taking the difference between the unweighted and the weighted distribution into account.
Both pileup and boson \pt reweighting uncertainties are constructed to affect only the shape and not the normalization of the distributions.
The finite size of the simulation samples limits the achievable precision of the results, the corresponding systematic uncertainty is modeled by applying bin-by-bin statistical uncertainties in the fit using the Barlow--Beeston ``lite'' approach~\cite{BARLOW1993219}.

The simulations rely on choices made in the theoretical description of the process.
Similar to the reweighting uncertainties, the uncertainties in the theoretical description are constructed to have only a shape but no normalization effect.
The uncertainty in the PDFs is estimated by evaluating the set of Hessian uncertainties in the NNPDF 3.1 set, treating each of the 100 variations as an independent nuisance parameter in the measurement.
An additional uncertainty in the PDFs is added to account for the choice of the value of the strong coupling at the \PZ boson mass $\alpS(m_{\PZ}) = 0.118 \pm 0.002$ at which the PDFs are evaluated.
The uncertainties in the renormalization and factorization scales are accounted for by varying these scales up and down by a factor of 2 or 1/2 (omitting the extreme case where both scales are varied in opposite directions).
The uncertainty from the choice of \alpS for the initial- and final-state radiation (ISR, FSR) uncertainty of the parton shower is estimated using up and down variations of the ISR and FSR parton shower scales with a factor of 2.

The dominating source of systematic uncertainty is the uncertainty in the integrated luminosity, which is measured with a precision of 1.4\% after a calibration with a method based on van der Meer scans and including systematic effects in the integration of detector rates~\cite{CMS:2021xjt,CMS-PAS-LUM-22-001}.

The sizes of the systematic uncertainties in percentage of the fit results for the fiducial cross section measurement are summarized in Table~\ref{tab:syst_xs} and for their ratios in Table~\ref{tab:syst_ratio}. For the \WpoverWm ratio, the leading uncertainties cancel to a large extent, as a result of strong positive correlations between the systematic uncertainties in these two regions. For the \WoverZ ratios, the magnitude of the leading uncertainties remains comparable to those in the \PW regions. For the muon efficiency, this follows from the similarity of the muon properties from a \PW or \PZ boson decay, resulting in an effective cancellation of the systematic impact of one muon. Uncertainties with asymmetric uncertainty magnitudes for the different cross sections do not change significantly. Some of the smaller uncertainties are negatively correlated and therefore increase for the cross section ratios, e.g., the uncertainty in the muon momentum resolution correction, which has a strong effect on the dimuon mass shape but a negligible effect on the transverse mass shape, where the resolution is dominated by the missing transverse momentum. In the fit, this strong effect in the dimuon mass shape is buffered through variations of further systematic uncertainties, leading to the different behavior in the two regions and therefore resulting in negative correlation.

\begin{table}[!ht]
\centering
\topcaption{Post-fit uncertainties in percent for the fiducial cross section measurement. For completeness, also the integrated luminosity and statistical uncertainty are given.}
\label{tab:syst_xs}
\begin{tabular}{lcccc}
    \hline
    \multirow{2}{*}{Uncertainty source} & \multicolumn{4}{c}{Uncertainty in $\sigfid\BR$ (in \%) for} \\[2pt]
    & \makebox[1.2cm][c]{\PWp\!\!} & \makebox[1.2cm][c]{\PWm\!\!} & \makebox[1.2cm][c]{\PWpm\!\!} & \makebox[1.2cm][c]{\PZ} \\
    \hline
    Muon efficiency & 0.28 & 0.29 & 0.29 & 0.40 \\
    Finite size of MC samples (bin-by-bin) & 0.27 & 0.27 & 0.25 & 0.08 \\
    QCD background & 0.53 & 0.49 & 0.49 & 0.07 \\
    PDF, scales, and parton shower & 0.25 & 0.25 & 0.25 & 0.06 \\
    Muon momentum correction & 0.01 & 0.02 & 0.01 & 0.03 \\
    Recoil correction & 0.09 & 0.08 & 0.08 & 0.02 \\
    EWK background normalization & 0.05 & 0.05 & 0.05 & 0.02 \\
    \PZ boson \pt correction & 0.03 & 0.04 & 0.03 & 0.01 \\
    \ttbar background normalization & 0.01 & 0.03 & 0.02 & 0.01 \\
    Pileup & 0.01 & 0.02 & \makebox[0pt][r]{$<$}0.01 & \makebox[0pt][r]{$<$}0.01 \\[\cmsTabSkip]
    Total & 0.68 & 0.66 & 0.65 & 0.42 \\[\cmsTabSkip]
    Integrated luminosity & 1.4\phantom{0} & 1.4\phantom{0} & 1.4\phantom{0} & 1.4\phantom{0} \\[\cmsTabSkip]
    Statistical uncertainty & 0.03 & 0.03 & 0.02 & 0.06 \\
    \hline
\end{tabular}
\end{table}

\begin{table}[!ht]
\centering
\topcaption{Post-fit uncertainties in percent for the fiducial cross section ratio measurement. For completeness, also the statistical uncertainty is given.}
\label{tab:syst_ratio}
\begin{tabular}{lcccc}
    \hline
    \multirow{2}{*}{Uncertainty source} & \multicolumn{4}{c}{Uncertainty in \Rfid (in \%) for} \\[2pt]
    & \WpoverZ & \WmoverZ & \WpmoverZ & \WpoverWm \\
    \hline
    Muon efficiency & 0.27 & 0.29 & 0.28 & 0.02 \\
    Finite size of MC samples (bin-by-bin) & 0.26 & 0.26 & 0.25 & 0.19 \\
    QCD background & 0.52 & 0.47 & 0.48 & 0.29 \\
    PDF, scales, and parton shower & 0.24 & 0.24 & 0.24 & 0.04 \\
    Muon momentum correction & 0.04 & 0.05 & 0.04 & 0.01 \\
    Recoil correction & 0.10 & 0.09 & 0.09 & 0.04 \\
    EWK background normalization & 0.07 & 0.07 & 0.07 & \makebox[0pt][r]{$<$}0.01 \\
    \PZ boson \pt correction & 0.04 & 0.05 & 0.05 & 0.01 \\
    \ttbar background normalization & 0.02 & 0.04 & 0.03 & 0.02 \\
    Pileup & 0.01 & 0.02 & \makebox[0pt][r]{$<$}0.01 & 0.03 \\[\cmsTabSkip]
    Total & 0.68 & 0.66 & 0.64 & 0.38 \\[\cmsTabSkip]
    Statistical uncertainty & 0.07 & 0.07 & 0.06 & 0.04 \\
    \hline
\end{tabular}
\end{table}

\section{Results}

\begin{figure}[!p]
\centering
\includegraphics[width=0.49\textwidth]{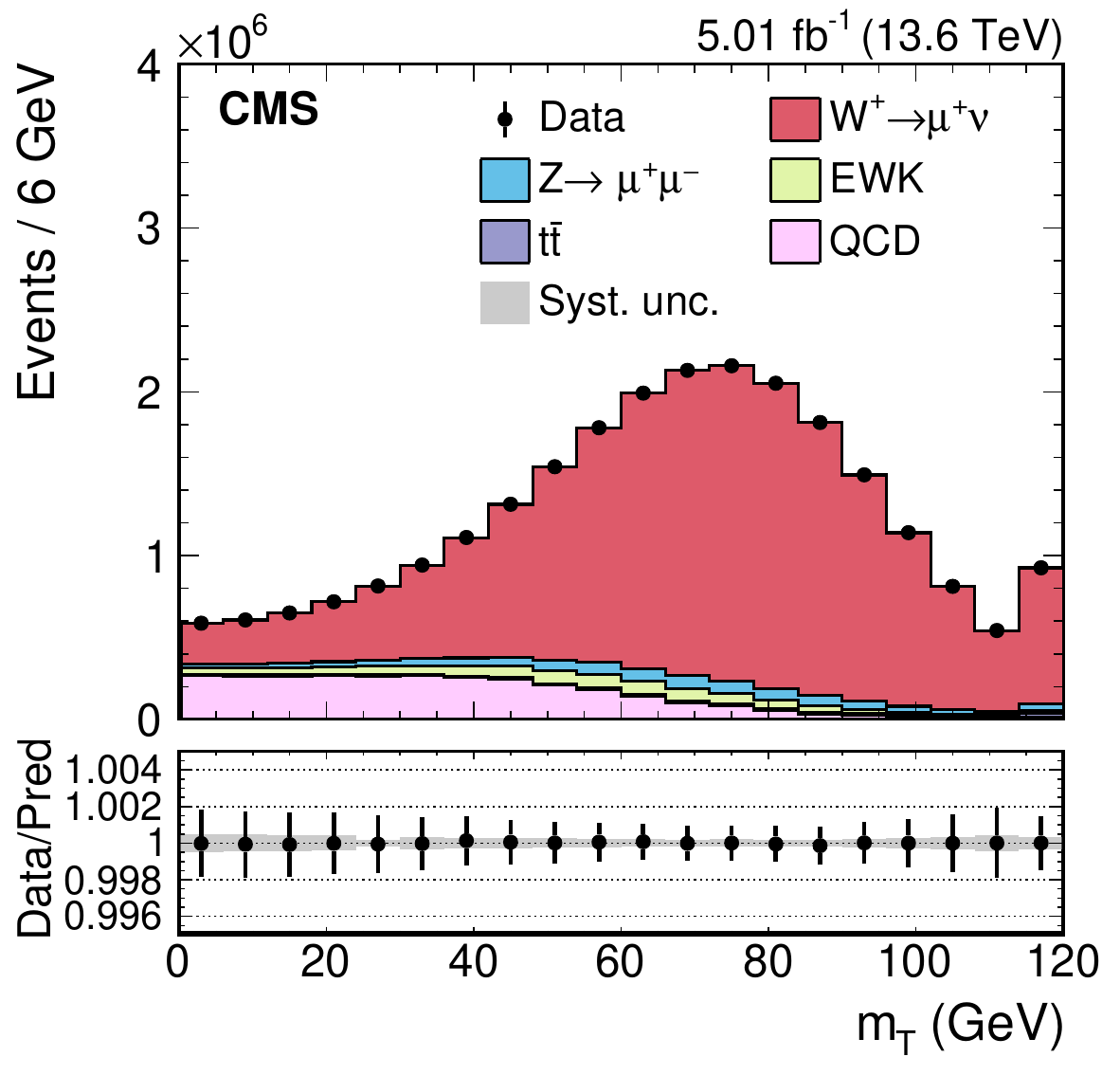}
\includegraphics[width=0.49\textwidth]{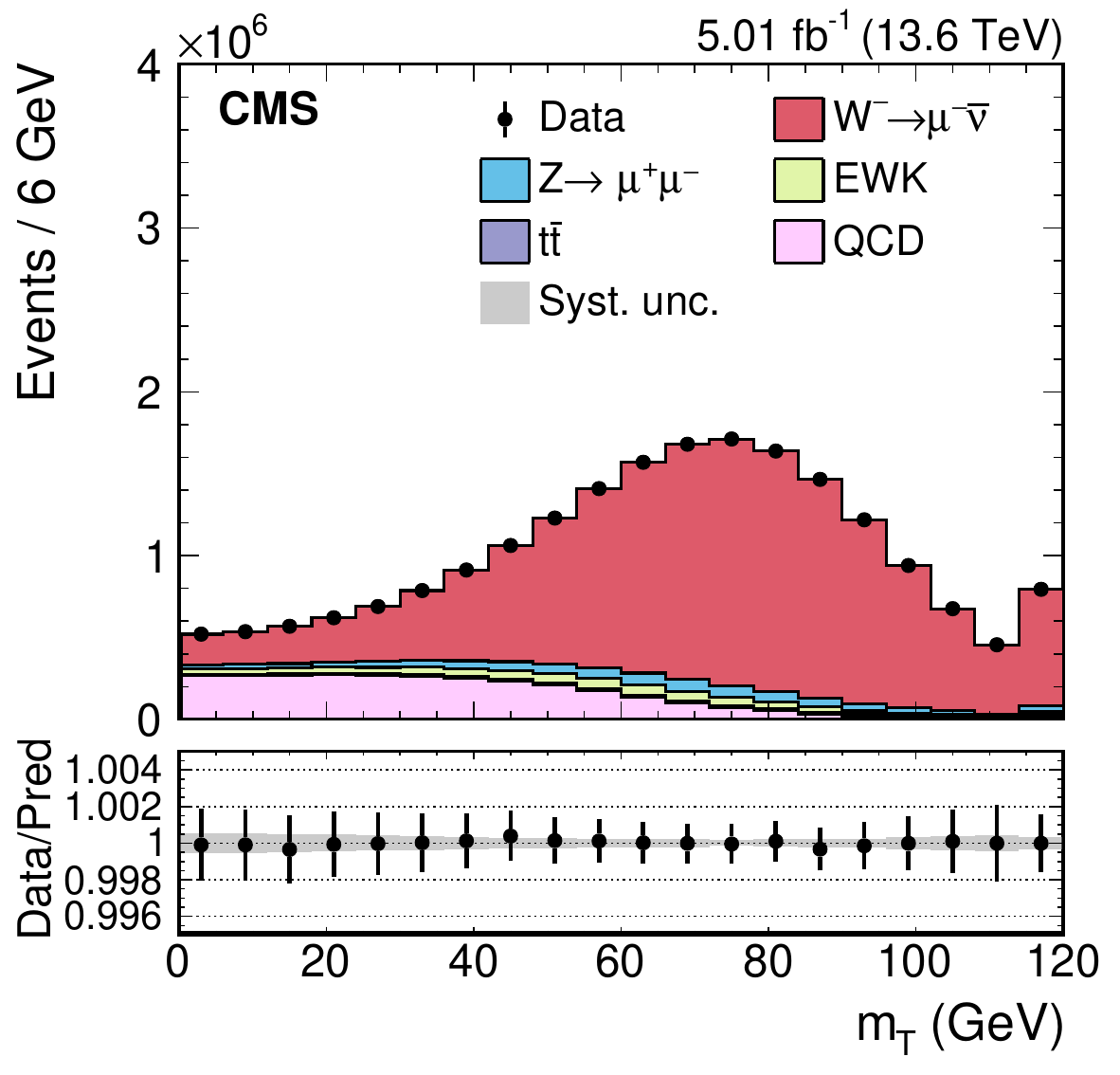}
\includegraphics[width=0.49\textwidth]{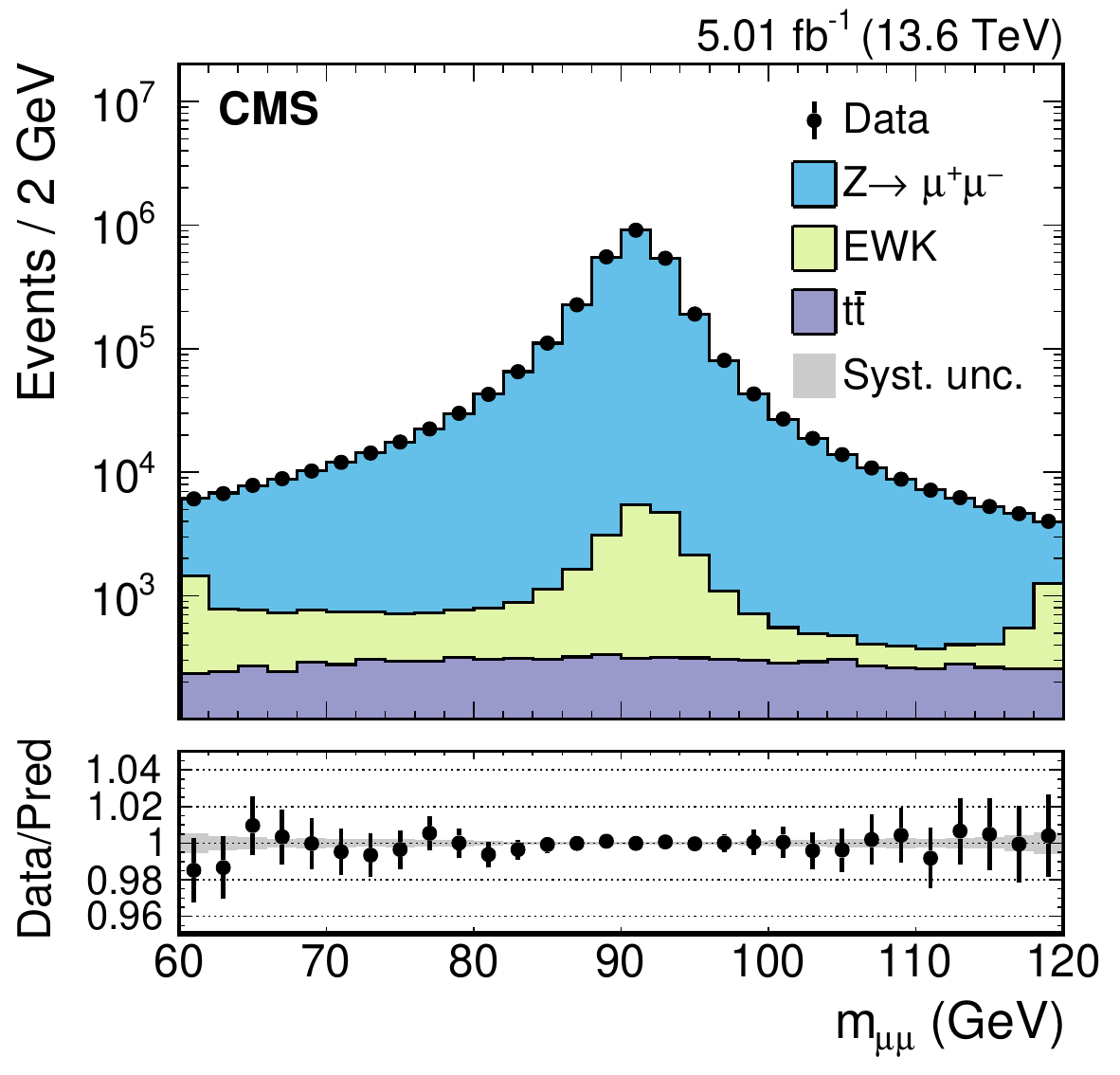}
\caption{The post-fit distributions of (upper left) $\PGmp\PGnGm$, (upper right) $\PGmm\PAGnGm$, and (lower) \MM signal regions. The lower panel in each plot shows the ratio of the number of events observed in data to that of the signal and background predictions. Overflow entries are included in the rightmost bin for the two upper plots. The vertical uncertainty bars on the data represent the statistical uncertainty.}
\label{fig:postfit_muon}
\end{figure}

Figure~\ref{fig:postfit_muon} shows the results of the fit to the \mT and \mmumu distributions in the single- and double-muon signal regions, respectively.
Good agreement between the model, which contains high degrees of freedom for the QCD multijet template, and the data is observed.
The expected event yields for the SM expectations of the different processes are shown in Table~\ref{tab:yield_pre} and the observed event yields are shown in Table~\ref{tab:yield_post}.

\begin{table}[!ht]
\centering
\topcaption{Pre-fit event yields in the fiducial region. A dash indicates that the corresponding contribution is found to be negligible in this signal region.}
\label{tab:yield_pre}
\renewcommand{\arraystretch}{1.1}
\begin{tabular}{lrrcrc}
    \hline
    Process & \PGmp region & \PGmm region & \multicolumn{3}{c}{\MM region} \\
    \hline
    \PWpm & 20\,035\,607 & 15\,591\,566 && \multicolumn{1}{c}{\NA} \\
    \PZ & 958\,372 & 896\,734 && 2\,944\,611 \\
    EWK: & (989\,729\makebox[0pt][l]{)} & (801\,381\makebox[0pt][l]{)} && (27\,526\makebox[0pt][l]{)} \\
    \qquad \PW ($\to \PGt\PGnGt$) & 608\,711 & 498\,487 && \multicolumn{1}{c}{\NA} \\
    \qquad \PZ ($\to \PGt\PGt$) & 119\,263 & 114\,194 && 2\,177 \\
    \qquad \PW (nonfiducial) & 121\,304 & 72\,035 && \multicolumn{1}{c}{\NA} \\
    \qquad \PZ (nonfiducial) & 9\,000 & 8\,789 && 16\,165 \\
    \qquad {\VV} & 45\,062 & 42\,333 && 6\,500 \\
    \qquad VBF \PV & 21\,284 & 16\,522 && 1\,680 \\
    \qquad Single top quark & 65\,106 & 49\,021 && 1\,004 \\
    \ttbar & 215\,409 & 216\,695 && 9\,278 \\
    \hline
\end{tabular}
\end{table}

\begin{table}[!ht]
\centering
\topcaption{Post-fit event yields in the fiducial region. The post-fit uncertainties include only statistical and systematic uncertainties, but not the uncertainty in the luminosity. The individual uncertainties in the event yields for a given process are derived by taking the full covariance matrix into account. A dash indicates that the corresponding contribution is found to be negligible in this signal region.}
\label{tab:yield_post}
\renewcommand{\arraystretch}{1.1}
\begin{tabular}{lr@{\,$\pm$\,}rr@{\,$\pm$\,}rr@{\,$\pm$\,}r}
    \hline
    Process & \multicolumn{2}{c}{\PGmp region} & \multicolumn{2}{c}{\PGmm region}  & \multicolumn{2}{c}{\MM region} \\
    \hline
    \PWpm             & 20\,013\,000 & 31\,000    & 15\,665\,000 & 22\,000    &  \multicolumn{2}{c}{\NA} \\
    \PZ               & 966\,700     & 2\,300     & 905\,000     & 2\,300     & 2\,962\,980 & 970 \\
    EWK               & 966\,000     & 14\,000    & 782\,000     & 11\,000    & 26\,560   & 460 \\
    \ttbar            & 200\,800     & 3\,400     & 202\,100     & 3\,400     & 8\,630      & 130 \\
    QCD multijet      & 2\,973\,000  & 33\,000    & 2\,920\,000  & 24\,000    & \multicolumn{2}{c}{\NA}  \\[\cmsTabSkip]
    Total             & 25\,119\,600    &   1\,500 & 20\,475\,200 &  1\,300&  2\,998\,160 &   850 \\[\cmsTabSkip]
    Data              & \multicolumn{2}{c}{25\,119\,622}  & \multicolumn{2}{c}{20\,475\,186}   &   \multicolumn{2}{c}{2\,998\,125}   \\
    \hline
\end{tabular}
\end{table}

\begin{figure}[!p]
   \centering
   \includegraphics[width=0.49\textwidth]{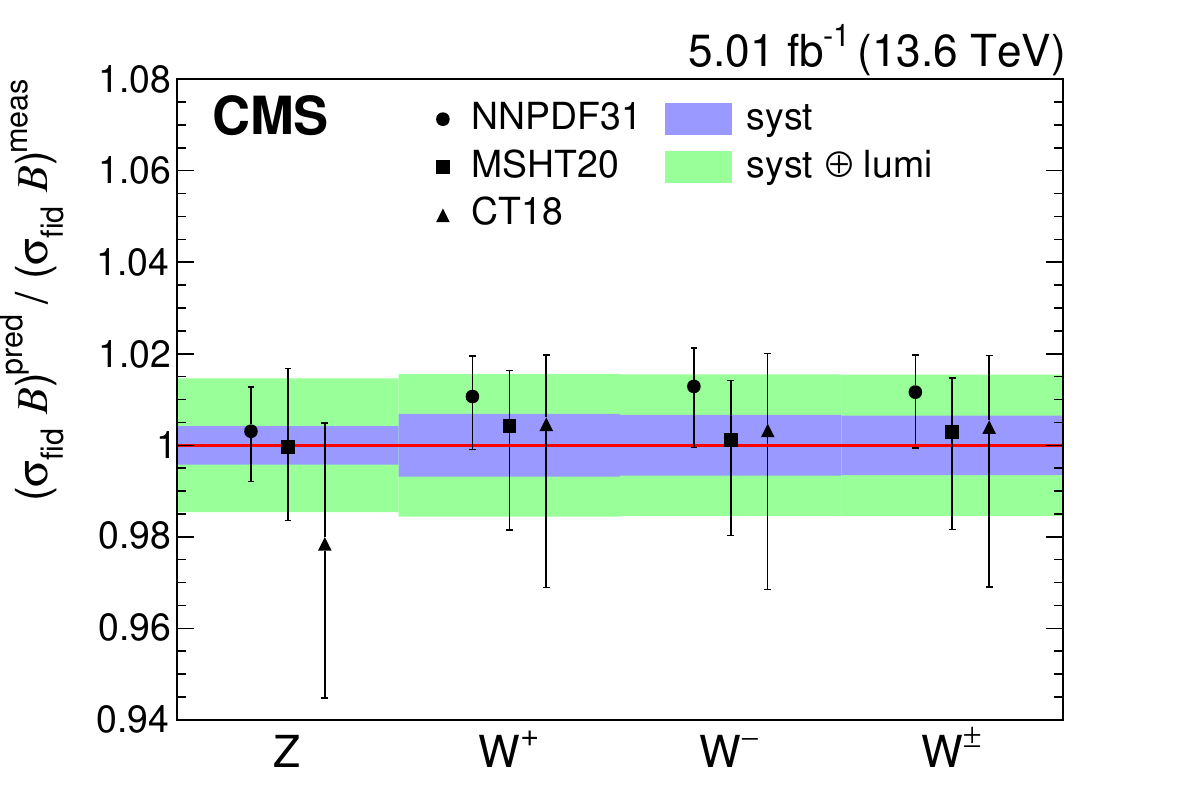}
   \includegraphics[width=0.49\textwidth]{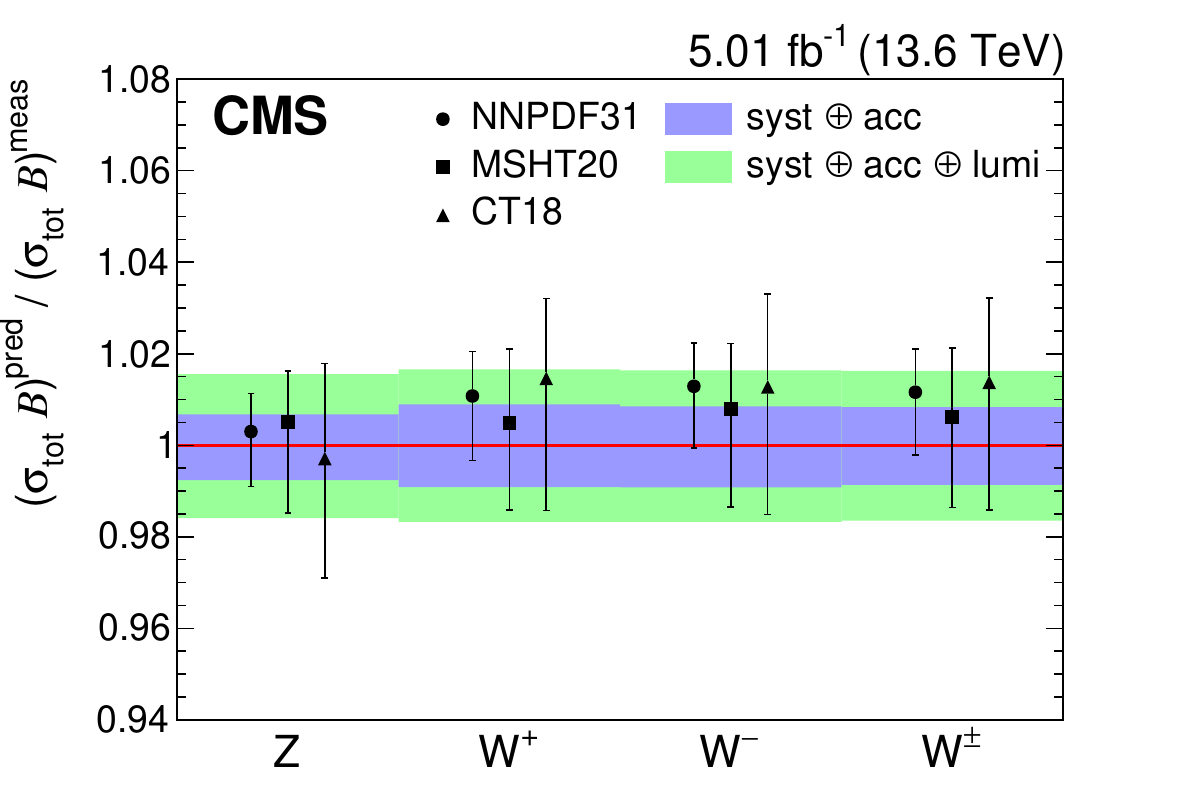}
   \caption{Comparison of measured product of fiducial (left) and total (right) cross sections and branching fractions with different theoretical predictions at NNLO+NNLL QCD accuracy. The vertical uncertainty bars on the markers represent the total uncertainty of the theoretical prediction.}
   \label{fig:xsec_comp}
\end{figure}

\begin{figure}[!p]
   \centering
   \includegraphics[width=0.49\textwidth]{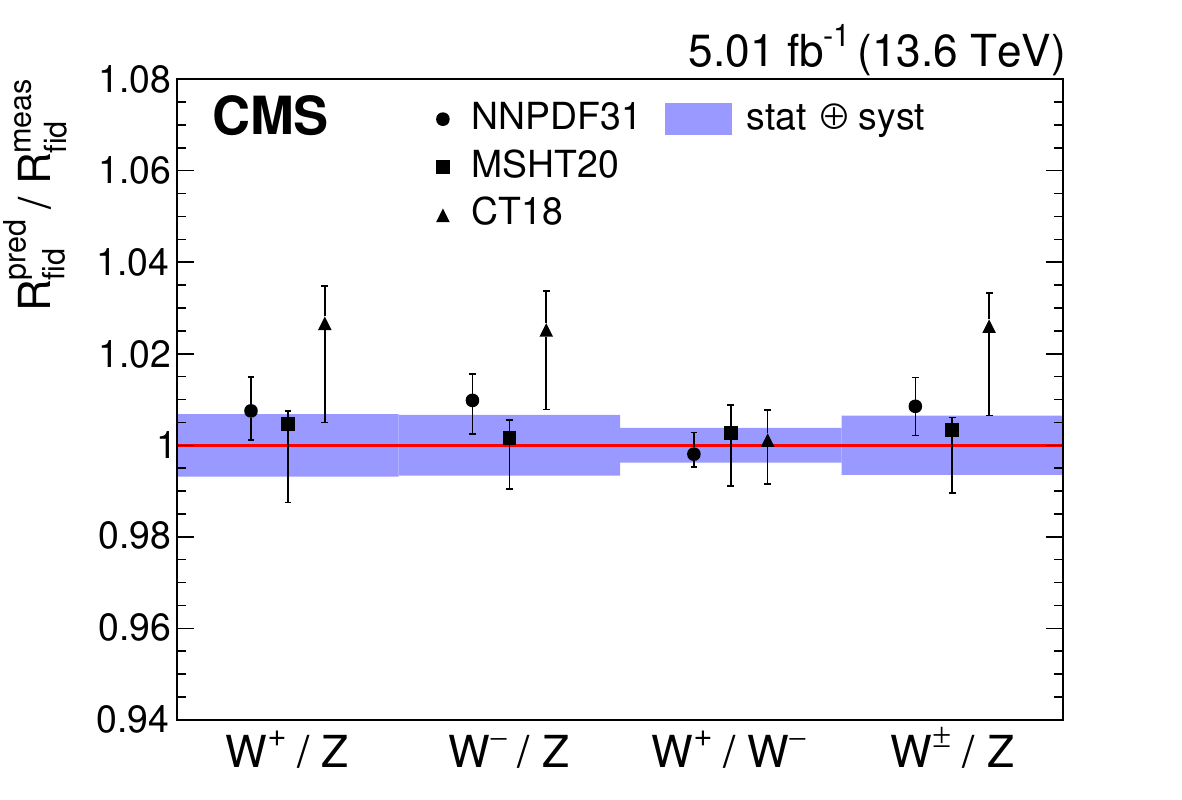}
   \includegraphics[width=0.49\textwidth]{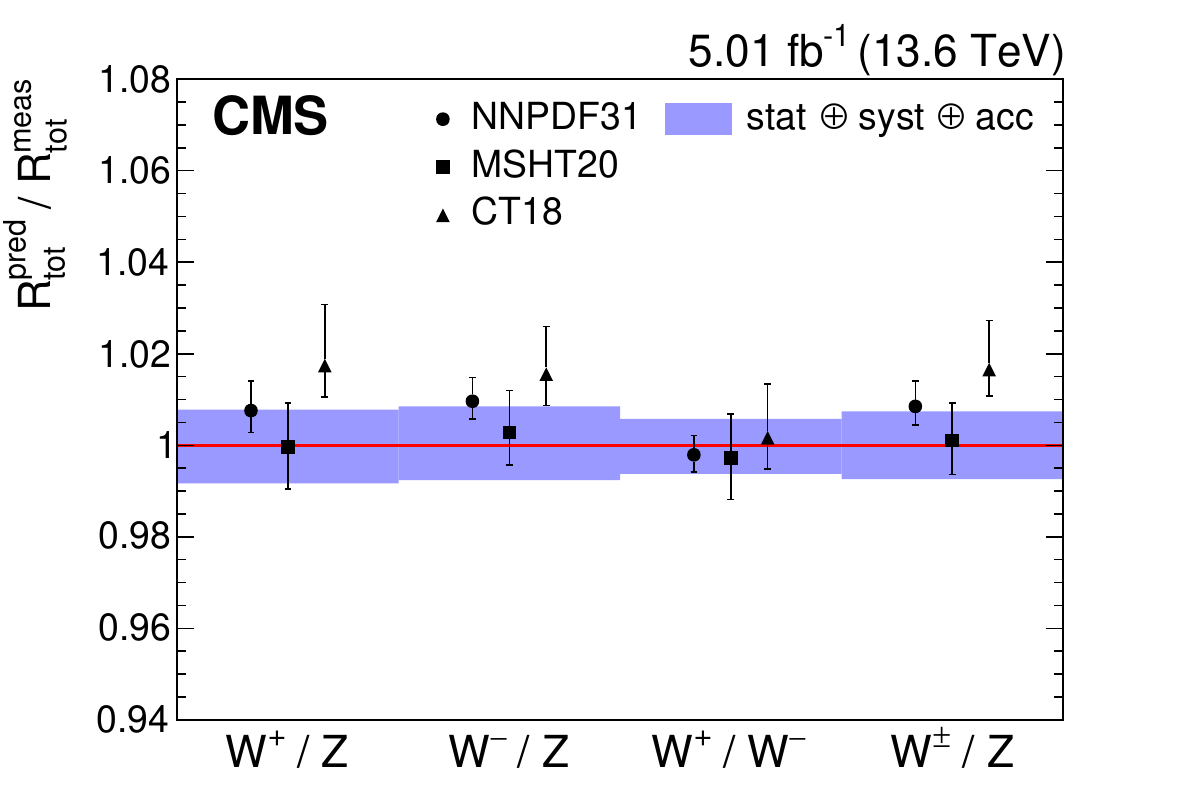}
   \caption{Comparison of measured fiducial (left) and total (right) ratios with different theoretical predictions at NNLO+NNLL QCD accuracy. The vertical uncertainty bars on the markers represent the total uncertainty of the theoretical prediction.}
   \label{fig:ratio_comp}
\end{figure}

The results for the product of the fiducial and total cross sections of inclusive \PWp, \PWm, \PWpm, and \PZ boson production and the branching fraction for the muon decay channels are summarized in Table~\ref{tab:resultsxsec}, along with the corresponding theoretical acceptance predictions.
The measured ratios of cross sections and their predicted acceptance values are given in Table~\ref{tab:resultsratios}.
Theoretical predictions of the product of the cross sections and branching fraction for the muonic decay are computed at NNLO+NNLL QCD accuracy in \DYTurbo using the NNPDF 3.1~\cite{NNPDF:2017mvq}, CT18~\cite{Hou:2019efy}, and MSHT20~\cite{Bailey:2020ooq} PDF sets.
Figure~\ref{fig:xsec_comp} shows comparisons between these predictions and the measured values.
The corresponding ratios of the cross sections are shown in Fig.~\ref{fig:ratio_comp}.
As the NNPDF 3.1 PDF set is used for the extrapolation of the fiducial result into the total phase space, the corresponding ratio of theoretical to measured values is consistent between the plots for the fiducial and total phase space.
For the other PDF sets, however, the ratio is not necessarily the same between the two phase spaces, as the acceptance of the corresponding PDF set and the one of the NNPDF 3.1 set used for the extrapolation of the measured value into the total phase space may differ.

\begin{table}[!p]
\centering
\topcaption{Results for the product of the fiducial and total inclusive cross sections and branching fractions measurements. For the measured values the quoted uncertainty represents the systematic uncertainty, while the statistical uncertainty is negligible. For the acceptance predictions, as explained in Section~\ref{sec:strategy}, the first uncertainty is the PDF uncertainty, the second is the scale uncertainty, and the third is the integration uncertainty of the calculation.}
\label{tab:resultsxsec}
\renewcommand{\arraystretch}{1.3}
\begin{tabular}{ll@{\,$\pm$\,}l@{\,$\pm$\,}ll@{\,$\pm$\,}l@{\,}l@{\,$\pm$\,}ll@{\,$\pm$\,}l@{\,$\pm$\,}l@{\,}l}
    \hline
    \multirow{2}{*}{Process} & \multicolumn{3}{c}{$\sigfid\BR\pm\systns$} & \multicolumn{4}{c}{Acceptance${}\pm\PDFns$} & \multicolumn{4}{c}{$\sigtot\BR\pm\systns\pm\lumns$} \\[-3pt]
    & \multicolumn{3}{c}{$\pm\lum$ in nb} & \multicolumn{4}{c}{$\pm\scale\pm\intns$ in \%} & \multicolumn{4}{c}{$\pm\acc$ in nb} \\
    \hline
    \PWp  & 5.428 & 0.037 & 0.076 & 45.48 & 0.24 & ${}^{+0.11}_{-0.07}$ & 0.07 & 11.93 & 0.08 & 0.17 & ${}^{+0.07}_{-0.07}$ \\
    \PWm  & 4.167 & 0.028 & 0.058 & 47.06 & 0.22 & ${}^{+0.19}_{-0.10}$ & 0.07 & \phantom{0}8.86 & 0.06 & 0.12 & ${}^{+0.05}_{-0.06}$ \\
    \PWpm & 9.60 & 0.06 & 0.13 & 46.15 & 0.22 & ${}^{+0.14}_{-0.08}$  & 0.05& 20.79 & 0.14 & 0.29 & ${}^{+0.11}_{-0.12}$ \\
    \PZ   & 0.764 & 0.003 & 0.011 & 37.80 & 0.19 & ${}^{+0.14}_{-0.04}$ & 0.04 & \phantom{0}2.021 & 0.009 & 0.028 & ${}^{+0.011}_{-0.013}$ \\
    \hline
\end{tabular}
\end{table}

\begin{table}[!p]
\centering
\topcaption{Ratios of the measured product of the fiducial and total inclusive cross sections and branching fractions along with the corresponding acceptance predictions. Since some contributions of the systematic uncertainty, most prominently the luminosity uncertainty, cancel out in the ratios, the statistical component becomes relevant. For the acceptance predictions, as explained in Section~\ref{sec:strategy}, the first uncertainty is the PDF uncertainty, the second is the scale uncertainty, and the third is the integration uncertainty of the calculation.}
\label{tab:resultsratios}
\renewcommand{\arraystretch}{1.3}
\begin{tabular}{lr@{\,$\pm$\,}l@{\,$\pm$\,}ll@{\,$\pm$\,}l@{\,}l@{\,$\pm$\,}lr@{\,$\pm$\,}l@{\,$\pm$\,}l@{\,}l}
    \hline
    \multirow{2}{*}{Process} & \multicolumn{3}{c}{\multirow{2}{*}{$\Rfid\pm\statns\pm\systns$}} & \multicolumn{4}{c}{Acceptance ratio} & \multicolumn{4}{c}{$\Rtot\pm\statns\pm\systns$} \\[-3pt]
    & \multicolumn{3}{c}{} & \multicolumn{4}{c}{$\pm\PDF\pm\scalens\pm\intns$} & \multicolumn{4}{c}{$\pm\acc$} \\
    \hline
    \WpoverZ & 7.105 & 0.005 & 0.048 & 1.203 & 0.003 & $^{+0.004}_{-0.002}$ & 0.002 & 5.906 & 0.004 & 0.040 & $^{+0.022}_{-0.028}$ \\
    \WmoverZ & 5.455 & 0.004 & 0.036 & 1.245 & 0.004 & $^{+0.001}_{-0.005}$ & 0.002 & 4.382 & 0.003 & 0.029 & $^{+0.023}_{-0.016}$ \\
    \WpmoverZ & 12.559 & 0.008 & 0.081 & 1.221 & 0.003 & $^{+0.003}_{-0.003}$ & 0.002 & 10.288 & 0.007 & 0.066 & $^{+0.037}_{-0.037}$ \\
    \WpoverWm & 1.303 & 0.001 & 0.005 & 0.966 & 0.003 & $^{+0.003}_{-0.002}$ & 0.002 & 1.348 & 0.001 & 0.005 & $^{+0.006}_{-0.007}$ \\
    \hline
\end{tabular}
\end{table}

\begin{figure}[!p]
\centering
\includegraphics[width=0.7\textwidth]{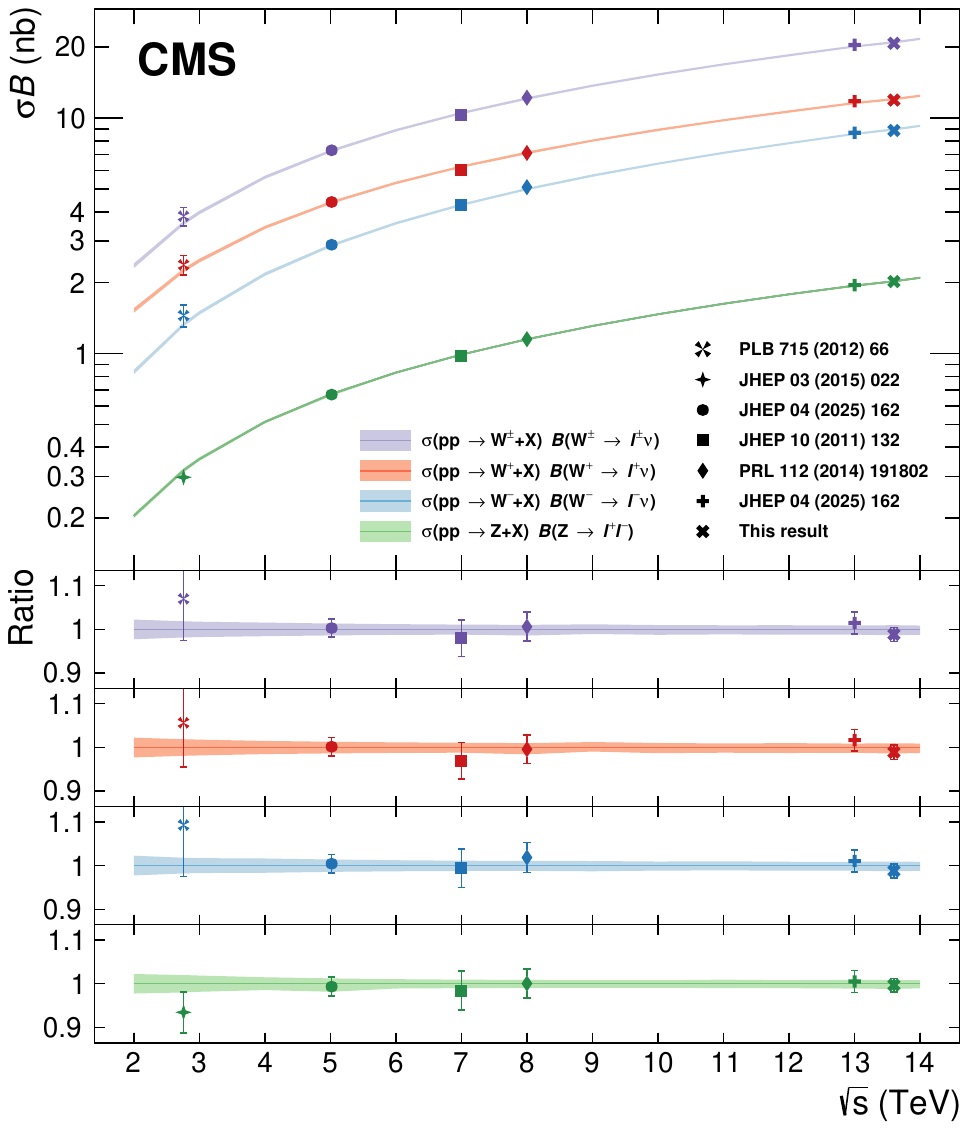}
\caption{Comparison of measured products of total cross section and branching fractions for \PW and \PZ boson production at different center-of-mass energies with the corresponding theoretical prediction at NNLO+NNLL QCD accuracy obtained from \DYTurbo. The uncertainties in the theoretical prediction include variations of the renormalization and factorization scales, as well as the PDF uncertainty evaluated with the NNPDF 3.1 set. The vertical uncertainty bars on the markers represent the total uncertainty of the measurement.}
\label{fig:xsecplot_z}
\end{figure}

A comparison of measured products of total cross section and branching fraction at other center-of-mass energies with the corresponding theoretical prediction is shown in Fig.~\ref{fig:xsecplot_z}.

\section{Summary}

Measurements of fiducial and total inclusive \PW and \PZ boson production cross sections multiplied by the $\PW\to\PGm\PGnGm$ and \ZtoMM branching fractions, respectively, in proton-proton collisions at 13.6\TeV are presented.
Muon final states are studied in data samples collected with the CMS detector corresponding to an integrated luminosity of $5.01\pm0.07\fbinv$.
The measured \PZ boson cross section is defined as the inclusive Drell--Yan \DY production of muon pairs in the invariant mass range of 60--120\GeV, where the production of a \PZ boson is the dominant contribution.
The results for the product of total inclusive cross sections and branching fractions are:
\begin{equation*}\begin{aligned}
    \sigma(\pp\to\PWp + \PX)\BR(\PWp\!\to\PGmp\PGnGm) &= 11.93\phantom{0} \pm 0.08\phantom{0} \syst \pm 0.17\phantom{0} \lum\,^{+0.07\phantom{0}}_{-0.07\phantom{0}} \acc \unit{nb},\\
    \sigma(\pp\to\PWm + \PX)\BR(\PWm\!\to\PGmm\PAGnGm) &= \phantom{0}8.86\phantom{0} \pm 0.06\phantom{0} \syst \pm 0.12\phantom{0} \lum\,^{+0.05\phantom{0}}_{-0.06\phantom{0}} \acc \unit{nb}, \\
    \sigma(\pp\to\PZ + \PX)\BR(\ZtoMM) &= \phantom{0}2.021 \pm 0.009 \syst \pm 0.028 \lum\,^{+0.011}_{-0.013} \acc \unit{nb},
\end{aligned}\end{equation*}
with the uncertainty in the integrated luminosity (lumi), the uncertainty in the acceptance of the fiducial volume, and the remaining systematic uncertainty (syst), whereas the statistical component  of the uncertainty is negligible.
This translates into cross section ratios of:
\begin{equation*}\begin{aligned}
    R^{\WpoverZ} &= 5.906 \pm 0.004 \stat \pm 0.040 \syst\,^{+0.022}_{-0.028} \acc, \\
    R^{\WmoverZ} &= 4.382 \pm 0.003 \stat \pm 0.029 \syst\,^{+0.023}_{-0.016} \acc, \\
    R^{\WpoverWm} &= 1.348 \pm 0.001 \stat \pm 0.005 \syst\,^{+0.006}_{-0.007} \acc,
\end{aligned}\end{equation*}
where the statistical component of the uncertainty (stat) is not negligible anymore due to the decreased size of the systematic contributions to the uncertainty in the cross section ratios.

All results are in agreement with theoretical calculations at next-to-next-to-leading order accuracy in quantum chromodynamics plus next-to-next-to-leading logarithmic \qt resummation, where \qt is the transverse momentum of the \PW or \PZ boson.
The high precision of the measurement highlights the excellent performance of the CMS detector following the long shutdown in 2018--2022, operating at the new center-of-mass energy of 13.6\TeV.
Moreover, the results underscore opportunities for refining the description of parton distribution functions and emphasize the potential value of dedicated differential measurements to enhance our understanding in the future.

\begin{acknowledgments}
    We congratulate our colleagues in the CERN accelerator departments for the excellent performance of the LHC and thank the technical and administrative staffs at CERN and at other CMS institutes for their contributions to the success of the CMS effort. In addition, we gratefully acknowledge the computing centers and personnel of the Worldwide LHC Computing Grid and other centers for delivering so effectively the computing infrastructure essential to our analyses. Finally, we acknowledge the enduring support for the construction and operation of the LHC, the CMS detector, and the supporting computing infrastructure provided by the following funding agencies: SC (Armenia), BMBWF and FWF (Austria); FNRS and FWO (Belgium); CNPq, CAPES, FAPERJ, FAPERGS, and FAPESP (Brazil); MES and BNSF (Bulgaria); CERN; CAS, MoST, and NSFC (China); MINCIENCIAS (Colombia); MSES and CSF (Croatia); RIF (Cyprus); SENESCYT (Ecuador); ERC PRG, RVTT3 and MoER TK202 (Estonia); Academy of Finland, MEC, and HIP (Finland); CEA and CNRS/IN2P3 (France); SRNSF (Georgia); BMBF, DFG, and HGF (Germany); GSRI (Greece); NKFIH (Hungary); DAE and DST (India); IPM (Iran); SFI (Ireland); INFN (Italy); MSIP and NRF (Republic of Korea); MES (Latvia); LMTLT (Lithuania); MOE and UM (Malaysia); BUAP, CINVESTAV, CONACYT, LNS, SEP, and UASLP-FAI (Mexico); MOS (Montenegro); MBIE (New Zealand); PAEC (Pakistan); MES and NSC (Poland); FCT (Portugal); MESTD (Serbia); MICIU/AEI and PCTI (Spain); MOSTR (Sri Lanka); Swiss Funding Agencies (Switzerland); MST (Taipei); MHESI and NSTDA (Thailand); TUBITAK and TENMAK (Turkey); NASU (Ukraine); STFC (United Kingdom); DOE and NSF (USA).

    \hyphenation{Rachada-pisek} Individuals have received support from the Marie-Curie program and the European Research Council and Horizon 2020 Grant, contract Nos.\ 675440, 724704, 752730, 758316, 765710, 824093, 101115353, 101002207, and COST Action CA16108 (European Union); the Leventis Foundation; the Alfred P.\ Sloan Foundation; the Alexander von Humboldt Foundation; the Science Committee, project no. 22rl-037 (Armenia); the Fonds pour la Formation \`a la Recherche dans l'Industrie et dans l'Agriculture (FRIA-Belgium); the Beijing Municipal Science \& Technology Commission, No. Z191100007219010 and Fundamental Research Funds for the Central Universities (China); the Ministry of Education, Youth and Sports (MEYS) of the Czech Republic; the Shota Rustaveli National Science Foundation, grant FR-22-985 (Georgia); the Deutsche Forschungsgemeinschaft (DFG), among others, under Germany's Excellence Strategy -- EXC 2121 ``Quantum Universe" -- 390833306, and under project number 400140256 - GRK2497; the Hellenic Foundation for Research and Innovation (HFRI), Project Number 2288 (Greece); the Hungarian Academy of Sciences, the New National Excellence Program - \'UNKP, the NKFIH research grants K 131991, K 133046, K 138136, K 143460, K 143477, K 146913, K 146914, K 147048, 2020-2.2.1-ED-2021-00181, TKP2021-NKTA-64, and 2021-4.1.2-NEMZ\_KI-2024-00036 (Hungary); the Council of Science and Industrial Research, India; ICSC -- National Research Center for High Performance Computing, Big Data and Quantum Computing and FAIR -- Future Artificial Intelligence Research, funded by the NextGenerationEU program (Italy); the Latvian Council of Science; the Ministry of Education and Science, project no. 2022/WK/14, and the National Science Center, contracts Opus 2021/41/B/ST2/01369 and 2021/43/B/ST2/01552 (Poland); the Funda\c{c}\~ao para a Ci\^encia e a Tecnologia, grant CEECIND/01334/2018 (Portugal); the National Priorities Research Program by Qatar National Research Fund; MICIU/AEI/10.13039/501100011033, ERDF/EU, "European Union NextGenerationEU/PRTR", and Programa Severo Ochoa del Principado de Asturias (Spain); the Chulalongkorn Academic into Its 2nd Century Project Advancement Project, and the National Science, Research and Innovation Fund via the Program Management Unit for Human Resources \& Institutional Development, Research and Innovation, grant B39G670016 (Thailand); the Kavli Foundation; the Nvidia Corporation; the SuperMicro Corporation; the Welch Foundation, contract C-1845; and the Weston Havens Foundation (USA).
\end{acknowledgments}

\bibliography{auto_generated}
\cleardoublepage \appendix\section{The CMS Collaboration \label{app:collab}}\begin{sloppypar}\hyphenpenalty=5000\widowpenalty=500\clubpenalty=5000
\cmsinstitute{Yerevan Physics Institute, Yerevan, Armenia}
{\tolerance=6000
V.~Chekhovsky, A.~Hayrapetyan, V.~Makarenko\cmsorcid{0000-0002-8406-8605}, A.~Tumasyan\cmsAuthorMark{1}\cmsorcid{0009-0000-0684-6742}
\par}
\cmsinstitute{Institut f\"{u}r Hochenergiephysik, Vienna, Austria}
{\tolerance=6000
W.~Adam\cmsorcid{0000-0001-9099-4341}, J.W.~Andrejkovic, L.~Benato\cmsorcid{0000-0001-5135-7489}, T.~Bergauer\cmsorcid{0000-0002-5786-0293}, S.~Chatterjee\cmsorcid{0000-0003-2660-0349}, K.~Damanakis\cmsorcid{0000-0001-5389-2872}, M.~Dragicevic\cmsorcid{0000-0003-1967-6783}, P.S.~Hussain\cmsorcid{0000-0002-4825-5278}, M.~Jeitler\cmsAuthorMark{2}\cmsorcid{0000-0002-5141-9560}, N.~Krammer\cmsorcid{0000-0002-0548-0985}, A.~Li\cmsorcid{0000-0002-4547-116X}, D.~Liko\cmsorcid{0000-0002-3380-473X}, I.~Mikulec\cmsorcid{0000-0003-0385-2746}, J.~Schieck\cmsAuthorMark{2}\cmsorcid{0000-0002-1058-8093}, R.~Sch\"{o}fbeck\cmsAuthorMark{2}\cmsorcid{0000-0002-2332-8784}, D.~Schwarz\cmsorcid{0000-0002-3821-7331}, M.~Sonawane\cmsorcid{0000-0003-0510-7010}, W.~Waltenberger\cmsorcid{0000-0002-6215-7228}, C.-E.~Wulz\cmsAuthorMark{2}\cmsorcid{0000-0001-9226-5812}
% ...
\par}
\vskip\cmsinstskip
\dag:~Deceased\\
$^{1}$Also at Yerevan State University, Yerevan, Armenia\\
$^{2}$Also at TU Wien, Vienna, Austria\\
$^{3}$Also at Ghent University, Ghent, Belgium\\
$^{4}$Also at Universidade do Estado do Rio de Janeiro, Rio de Janeiro, Brazil\\
$^{5}$Also at FACAMP - Faculdades de Campinas, Sao Paulo, Brazil\\
$^{6}$Also at Universidade Estadual de Campinas, Campinas, Brazil\\
$^{7}$Also at Federal University of Rio Grande do Sul, Porto Alegre, Brazil\\
$^{8}$Also at University of Chinese Academy of Sciences, Beijing, China\\
$^{9}$Also at China Center of Advanced Science and Technology, Beijing, China\\
$^{10}$Also at University of Chinese Academy of Sciences, Beijing, China\\
$^{11}$Also at China Spallation Neutron Source, Guangdong, China\\
$^{12}$Now at Henan Normal University, Xinxiang, China\\
$^{13}$Also at University of Shanghai for Science and Technology, Shanghai, China\\
$^{14}$Now at The University of Iowa, Iowa City, Iowa, USA\\
$^{15}$Also at an institute formerly covered by a cooperation agreement with CERN\\
$^{16}$Also at Helwan University, Cairo, Egypt\\
$^{17}$Now at Zewail City of Science and Technology, Zewail, Egypt\\
$^{18}$Now at British University in Egypt, Cairo, Egypt\\
$^{19}$Now at Cairo University, Cairo, Egypt\\
$^{20}$Also at Purdue University, West Lafayette, Indiana, USA\\
$^{21}$Also at Universit\'{e} de Haute Alsace, Mulhouse, France\\
$^{22}$Also at Istinye University, Istanbul, Turkey\\
$^{23}$Also at Tbilisi State University, Tbilisi, Georgia\\
$^{24}$Also at an international laboratory covered by a cooperation agreement with CERN\\
$^{25}$Also at The University of the State of Amazonas, Manaus, Brazil\\
$^{26}$Also at University of Hamburg, Hamburg, Germany\\
$^{27}$Also at RWTH Aachen University, III. Physikalisches Institut A, Aachen, Germany\\
$^{28}$Also at Bergische University Wuppertal (BUW), Wuppertal, Germany\\
$^{29}$Also at Brandenburg University of Technology, Cottbus, Germany\\
$^{30}$Also at Forschungszentrum J\"{u}lich, Juelich, Germany\\
$^{31}$Also at CERN, European Organization for Nuclear Research, Geneva, Switzerland\\
$^{32}$Also at HUN-REN ATOMKI - Institute of Nuclear Research, Debrecen, Hungary\\
$^{33}$Now at Universitatea Babes-Bolyai - Facultatea de Fizica, Cluj-Napoca, Romania\\
$^{34}$Also at MTA-ELTE Lend\"{u}let CMS Particle and Nuclear Physics Group, E\"{o}tv\"{o}s Lor\'{a}nd University, Budapest, Hungary\\
$^{35}$Also at HUN-REN Wigner Research Centre for Physics, Budapest, Hungary\\
$^{36}$Also at Physics Department, Faculty of Science, Assiut University, Assiut, Egypt\\
$^{37}$Also at Punjab Agricultural University, Ludhiana, India\\
$^{38}$Also at University of Visva-Bharati, Santiniketan, India\\
$^{39}$Also at Indian Institute of Science (IISc), Bangalore, India\\
$^{40}$Also at Amity University Uttar Pradesh, Noida, India\\
$^{41}$Also at IIT Bhubaneswar, Bhubaneswar, India\\
$^{42}$Also at Institute of Physics, Bhubaneswar, India\\
$^{43}$Also at University of Hyderabad, Hyderabad, India\\
$^{44}$Also at Deutsches Elektronen-Synchrotron, Hamburg, Germany\\
$^{45}$Also at Isfahan University of Technology, Isfahan, Iran\\
$^{46}$Also at Sharif University of Technology, Tehran, Iran\\
$^{47}$Also at Department of Physics, University of Science and Technology of Mazandaran, Behshahr, Iran\\
$^{48}$Also at Department of Physics, Faculty of Science, Arak University, ARAK, Iran\\
$^{49}$Also at Italian National Agency for New Technologies, Energy and Sustainable Economic Development, Bologna, Italy\\
$^{50}$Also at Centro Siciliano di Fisica Nucleare e di Struttura Della Materia, Catania, Italy\\
$^{51}$Also at Universit\`{a} degli Studi Guglielmo Marconi, Roma, Italy\\
$^{52}$Also at Scuola Superiore Meridionale, Universit\`{a} di Napoli 'Federico II', Napoli, Italy\\
$^{53}$Also at Fermi National Accelerator Laboratory, Batavia, Illinois, USA\\
$^{54}$Also at Lulea University of Technology, Lulea, Sweden\\
$^{55}$Also at Consiglio Nazionale delle Ricerche - Istituto Officina dei Materiali, Perugia, Italy\\
$^{56}$Also at Institut de Physique des 2 Infinis de Lyon (IP2I ), Villeurbanne, France\\
$^{57}$Also at Department of Applied Physics, Faculty of Science and Technology, Universiti Kebangsaan Malaysia, Bangi, Malaysia\\
$^{58}$Also at Consejo Nacional de Ciencia y Tecnolog\'{i}a, Mexico City, Mexico\\
$^{59}$Also at Trincomalee Campus, Eastern University, Sri Lanka, Nilaveli, Sri Lanka\\
$^{60}$Also at Saegis Campus, Nugegoda, Sri Lanka\\
$^{61}$Also at National and Kapodistrian University of Athens, Athens, Greece\\
$^{62}$Also at Ecole Polytechnique F\'{e}d\'{e}rale Lausanne, Lausanne, Switzerland\\
$^{63}$Also at Universit\"{a}t Z\"{u}rich, Zurich, Switzerland\\
$^{64}$Also at Stefan Meyer Institute for Subatomic Physics, Vienna, Austria\\
$^{65}$Also at Laboratoire d'Annecy-le-Vieux de Physique des Particules, IN2P3-CNRS, Annecy-le-Vieux, France\\
$^{66}$Also at Near East University, Research Center of Experimental Health Science, Mersin, Turkey\\
$^{67}$Also at Konya Technical University, Konya, Turkey\\
$^{68}$Also at Izmir Bakircay University, Izmir, Turkey\\
$^{69}$Also at Adiyaman University, Adiyaman, Turkey\\
$^{70}$Also at Bozok Universitetesi Rekt\"{o}rl\"{u}g\"{u}, Yozgat, Turkey\\
$^{71}$Also at Marmara University, Istanbul, Turkey\\
$^{72}$Also at Milli Savunma University, Istanbul, Turkey\\
$^{73}$Also at Kafkas University, Kars, Turkey\\
$^{74}$Now at Istanbul Okan University, Istanbul, Turkey\\
$^{75}$Also at Hacettepe University, Ankara, Turkey\\
$^{76}$Also at Erzincan Binali Yildirim University, Erzincan, Turkey\\
$^{77}$Also at Istanbul University -  Cerrahpasa, Faculty of Engineering, Istanbul, Turkey\\
$^{78}$Also at Yildiz Technical University, Istanbul, Turkey\\
$^{79}$Also at School of Physics and Astronomy, University of Southampton, Southampton, United Kingdom\\
$^{80}$Also at IPPP Durham University, Durham, United Kingdom\\
$^{81}$Also at Monash University, Faculty of Science, Clayton, Australia\\
$^{82}$Also at Universit\`{a} di Torino, Torino, Italy\\
$^{83}$Also at Bethel University, St. Paul, Minnesota, USA\\
$^{84}$Also at Karamano\u {g}lu Mehmetbey University, Karaman, Turkey\\
$^{85}$Also at California Institute of Technology, Pasadena, California, USA\\
$^{86}$Also at United States Naval Academy, Annapolis, Maryland, USA\\
$^{87}$Also at Ain Shams University, Cairo, Egypt\\
$^{88}$Also at Bingol University, Bingol, Turkey\\
$^{89}$Also at Georgian Technical University, Tbilisi, Georgia\\
$^{90}$Also at Sinop University, Sinop, Turkey\\
$^{91}$Also at Erciyes University, Kayseri, Turkey\\
$^{92}$Also at Horia Hulubei National Institute of Physics and Nuclear Engineering (IFIN-HH), Bucharest, Romania\\
$^{93}$Now at another institute formerly covered by a cooperation agreement with CERN\\
$^{94}$Also at Texas A\&M University at Qatar, Doha, Qatar\\
$^{95}$Also at Kyungpook National University, Daegu, Korea\\
$^{96}$Also at Yerevan Physics Institute, Yerevan, Armenia\\
$^{97}$Also at another international laboratory covered by a cooperation agreement with CERN\\
$^{98}$Also at Imperial College, London, United Kingdom\\
$^{99}$Also at Institute of Nuclear Physics of the Uzbekistan Academy of Sciences, Tashkent, Uzbekistan\\
$^{100}$Also at another institute formerly covered by a cooperation agreement with CERN\\
\end{sloppypar}
\end{document}